\documentclass[preprint]{aastex631}

\setcitestyle{notesep={ }}

\begin{document}

\title{Investigating Performance Trends of Simulated Real-time Solar Flare Predictions: The Impacts of Training Windows, Data Volumes, and the Solar Cycle}

\author[0000-0003-3493-9174]{Griffin T. Goodwin}

\author[0000-0002-4001-1295]{Viacheslav M. Sadykov}

\author[0000-0001-8078-6856]{Petrus C. Martens}
\affiliation{Physics \& Astronomy Department, Georgia State University, Atlanta, GA 30303, USA; \href{mailto:ggoodwin5@gsu.edu}{ggoodwin5@gsu.edu}}


\begin{abstract}
This study explores the behavior of machine learning-based flare forecasting models deployed in a simulated operational environment. Using Georgia State University's Space Weather Analytics for Solar Flares benchmark dataset \citep{DV_2020,angryk2020multivariate}, we examine the impacts of training methodology and the solar cycle on decision tree, support vector machine, and multilayer perceptron performance. We implement our classifiers using three temporal training windows: stationary, rolling, and expanding. The stationary window trains models using a single set of data available before the first forecasting instance, which remains constant throughout the solar cycle. The rolling window trains models using data from a constant time interval before the forecasting instance, which moves with the solar cycle. Finally, the expanding window trains models using all available data before the forecasting instance. For each window, a number of input features (1, 5, 10, 25, 50, 120) and temporal sizes (5, 8, 11, 14, 17, 20 months) were tested. To our surprise, we found that for a 20-month window, skill scores were comparable regardless of the window type, feature count, and classifier selected. Furthermore, reducing the size of this window only marginally decreased stationary and rolling window performance. This implies that, given enough data, a stationary window can be chosen over other window types, eliminating the need for model retraining. Lastly, a moderately strong positive correlation was found to exist between a model's false positive rate and the solar X-ray background flux. This suggests that the solar cycle phase has a considerable influence on forecasting.
\end{abstract}



\section{Introduction} \label{sec:intro}
Due to humanity’s growing technological advancements over the past century, solar eruptive events have emerged as a significant threat to society and its infrastructure. Electromagnetic radiation, solar energetic particles, and coronal mass ejections produced during solar flares have the potential to interfere with radio communications, GPS, and power grids \citep{natras2019strong,hudson2021carrington}, which are crucial components to our everyday lives. Furthermore, these events pose considerable health risks to humans, particularly astronauts who are not shielded by Earth’s magnetosphere and may receive increased doses of radiation.  Considering these effects, the need for robust forecasting models that provide accurate and timely predictions of solar flares has become increasingly important. Traditional forecasting methods, such as those used by the National Oceanic and Atmospheric Administration (NOAA), have long relied on a blend of statistical analyses and human intuition \citep{crown2012noaaforecast}. However, given the advancements of artificial intelligence in recent years, there has been a gradual shift towards utilizing machine learning (ML) to automate and improve current forecasting capabilities. ML centers on training computers to make predictions on unseen data, given their previously acquired knowledge of some available dataset \citep{florios2018forecasting}. For flares, this can be physics-based parameters of active region (AR) vector magnetograms, extreme ultraviolet images of ARs, or even sunspot properties and McIntosh classifications \citep{bobra2015solar,2018ApJ...858..113N,li2007support}. Since its initial application to space weather in the early 1990s \citep{challengesML}, ML has grown significantly, showing great promise within the community. However, despite this success, several notable issues continue to limit its implementation in operational forecasting: 
\begin{enumerate}
    \item ML models are commonly trained and tested using a random set of flaring and non-flaring data, which is not necessarily consistent with real-time forecasting. In an operational setting, predictions must be based solely on data available prior to the forecasted event. This raises the question: How do ML classifiers perform when utilizing chronological training and testing partitions? \citet{sadykov2017PIL,2018ApJ...858..113N,leka2019comparison} have considered this idea through static training and testing windows, however, to the best of our knowledge, no studies have attempted to implement a dynamic temporal training strategy to improve operational forecasts.
    \item Complicated ML algorithms are often considered black boxes, providing little insight into their predictive reasoning \citep{challengesML}. This makes it challenging for forecasters to rely on them confidently. Thankfully, relatively basic models exist that provide easily interpretable predictions. However, there is no guarantee that these models perform as well as their more complex counterparts. A previous study from \cite{2023A&A...674A.159D} found that, for flare forecasting, ML models of different complexities were quite comparable, but a similar investigation has yet to be done for a real-time forecasting environment.
    \item ML models are frequently trained on all available data to maximize performance. However, this can result in a time-consuming training and hyperparameter optimization phase, which is not ideal for real-time forecasting. A middle ground between performance and run time likely exists, but the amount of data necessary to generate effective flare forecasts is currently poorly understood.
    \item The performance of ML-based flare forecasting models is heavily influenced by the selection of training data. Previous studies have shown that skill scores may vary significantly when training on different parts of the solar cycle \citep{wang2020predicting}. It is unclear whether these impacts can be mitigated through dynamic training windows.
\end{enumerate}

The goal of this work is to thoroughly examine each of these concerns. To address \textbf{Problem 1}, we deploy a training and testing methodology that simulates a real-time predictive environment. We accomplish this through three training windows we label as stationary, rolling, and expanding. For \textbf{Problem 2}, we apply our training methodology to three different ML models of increasing complexity: decision tree, support vector machine, and multilayer perceptron. We then explore how performance scales with the number of magnetogram features used in a prediction. To tackle \textbf{Problem 3}, we investigate the impact of data volume on performance by implementing different stationary and rolling window sizes. Finally, to handle \textbf{Problem 4}, we explore the relationship between classifier performance and the solar background soft X-ray (SXR) flux. We use this as a probe to investigate if the solar cycle has an effect on real-time forecasts, as well as how this potential dependency interacts with the dynamic training windows. We would like to emphasize that the main goal of this work is not necessarily to produce the best-performing classifier, but rather to exhaustively examine the problems we have mentioned above.

The remaining sections of this work are organized as follows: Section \ref{sec:data} details the data we use to construct our forecasts. Section \ref{sec:methods} describes the methodology used to tune, train, test, and analyze our ML models. Lastly, Section \ref{sec:results} \& \ref{sec:conc} highlight the results and conclusions of our study.

\section{Data}
\label{sec:data}
In this section, we provide a description of the three key datasets employed in this work. Section \ref{sec:SWANSF} provides an overview of the Space Weather Analytics for Solar Flares database, while Section \ref{sec:GOESX-Ray} briefly describes the data used to analyze the performance dependency on the solar cycle.
\subsection{Space Weather Analytics For Solar Flares (SWAN-SF)}
\label{sec:SWANSF}
Georgia State University’s Space Weather Analytics for Solar Flares (SWAN-SF) benchmark dataset \citep{angryk2020multivariate} is a comprehensive, ML-ready collection of multivariate time series samples extracted from ARs present during Solar Cycle 24 (May 2010 – August 2018). For each AR, twenty-four physics-based features \citep[see][Table 1]{angryk2020multivariate} are derived from photospheric vector magnetograms taken by the Solar Dynamics Observatory Helioseismic and Magnetic Imager \citep{2012SoPh..275..207S}. Throughout an AR's lifetime, time series data are sliced into temporally successive overlapping files (offset by 1-hour), each containing 12 hours’ worth of data, at a 12-minute cadence. Files are then labeled based on the strongest flaring event that occurs in the following 24 hours. We categorize flare strength using NOAA's logarithmic classification scale: A (weakest), B, C, M, and X (strongest). For this study, M and X-class flares are labeled as flaring events, considering they have the greatest potential for societal impacts. Weaker flares, in addition to flare quiet time series, are labeled as non-flaring events. 

In total, there are 331,185 AR multivariate time series files in SWAN-SF. Given this, along with the high dimensionality of each file, we decided to eliminate the contiguous temporal dimension of our data. This process not only allowed for easier integration with our proposed ML models, discussed in Section \ref{sec:MLC}, but significantly reduced training and testing times, which is an important aspect to consider when deploying models operationally. By extracting the summary statistics (mean, median, standard deviation, maximum, and minimum) of each magnetic field parameter, within the 12-hour window, all files were reduced to a single, point-in-time datum, with a dimension of 1 by 120. Any files containing columns with missing data were linearly interpolated before calculating their summary statistics, while files with empty columns were dropped altogether. To ensure that the data reflected a real-time forecasting scenario as much as possible, all ARs were retained, including those potentially prone to projection effects at radial distances greater than 70 degrees from the solar disk center. Ultimately, we were left with 330,169 data points: 6,234 flaring and 323,935 non-flaring. 

If the reader would like to learn more about the original data processing techniques for SWAN-SF, please refer to \cite{angryk2020multivariate}.

\subsection{Geostationary Operational Environmental Satellite (GOES) SXR Flux \& Hale Classifications}
\label{sec:GOESX-Ray}
To investigate potential correlations between a classifier's performance and the phase of the solar cycle, we utilized daily SXR flux data from GOES (1 - 8 \AA\space channel) provided by \cite{ali2024predicting}. A proxy for the solar SXR background flux was then determined by selecting the minimum X-ray flux measurement for each day during Solar Cycle 24. Additionally, we made use of AR Hale classifications provided by \cite{Marroquin_2023}, to study the frequency of complex ARs throughout the solar cycle. We then used these data in conjunction with each other to synthesize our results discussed in Section \ref{sec:SXR_Cor}. 

\section{Methodology} \label{sec:methods}
The structure of SWAN-SF frames this forecasting problem as a binary classification task, with the ultimate goal of determining whether an AR will produce a $\geq$M class solar flare within the next 24 hours. Previous work has shown that ML-based classifiers such as decision tree, logistic regression, random forest, support vector machine, and multilayer perceptron provide relatively reasonable forecasting performance when utilizing magnetogram feature sets \citep{yu2009short,yuan2010automated,bobra2015solar,florios2018forecasting}. In this particular study, we focus on the simulated real-time performance of three models: decision tree, support vector machine, and multilayer perception. This subset covers a wide gambit of complexities, ensuring we obtain robust results.

The following sections provide a basic overview of each model (Section \ref{sec:MLC}), the methodology for data preprocessing, feature selection, and hyperparameter tuning (Section \ref{sec:preprocess}), the design of each training window (Section \ref{sec:trainingwindows}), the performance metrics used to analyze our results (Section \ref{sec:performancemetrics}), and our approach for studying the solar cycle dependence (Section \ref{sec:dependenceonsolarcycle}).
\subsection{Machine Learning Classifiers}
\label{sec:MLC}
Decision trees (DT) are a simple, yet effective, ML algorithm for classification. Their foundation stems from a series of feature-based inequalities, which guide an input to a prediction. The overall structure of this model is hierarchical in nature, consisting of interconnected nodes, children, and leaves. Each node contains a test that compares a particular feature to some threshold value. These nodes are then split into child nodes, each with their own thresholds, further subdividing the tree. Eventually, enough splits are made to obtain a leaf, which determines the prediction for a given input. Mathematically, DTs are constructed by minimizing the impurity of successive splits of the training data. This can be considered analogous to determining the feature and decision boundary that best separates two labeled distributions \citep{kingsford2008decision,kotsiantis2013decision,2023A&A...674A.159D}. In this work, we focus on optimizing two key hyperparameters of our DT model: the tree depth and the number of training samples needed for a split/leaf node to occur. We also explore a variety of impurities (Gini and entropy), when splitting nodes (see \cite{kingsford2008decision,kotsiantis2013decision} for more details). 

In its simplest form, the support vector machine (SVM) algorithm identifies a multidimensional hyperplane in feature space that maximizes the separation between labels. This hyperplane is then used to make predictions on input data. Typically, more complex, non-linear structures are needed to separate labels adequately. Thus, kernels may be applied to map the feature space into a higher dimension. In this paper, we employ a radial basis function (RBF) kernel \citep{bobra2015solar}, which is influenced by two fundamental hyperparameters: $C$ and $\gamma$. $C$ is a penalty parameter for the misclassification of training data, where large values of $C$ result in overfitting and low values lead to underfitting. $\gamma$ on the other hand is the ``width" of the kernel. High values of $\gamma$ increase the complexity of the decision boundary, while smaller values generate a smoother division, resulting in performance similar to that of a linear boundary. 

Multilayer perceptrons (MLP) are a subclass of feed-forward neural networks containing a set of fully connected nodes (or neurons) across consecutive neuron layers. These nodes, which constitute the building blocks of the algorithm, possess a series of inputs, outputs, and weights that facilitate the creation of a non-linear decision boundary. All incoming data to a node is multiplied by its corresponding weight and summed together. The result is then transformed using an activation function and passed to the output, which connects to each node within the subsequent layer \citep{GARDNER19982627,2023A&A...674A.159D}. Typically, MLPs have three stages: a single input layer, a single output layer, and some arbitrarily large hidden layer sandwiched in between. For binary classification tasks, the input layer contains the same number of nodes as the size of the feature space, the output layer contains 2 nodes, and the hidden layers contain any number of nodes. In this study, after some trial and error, we settled on a 3-stage hidden layer, with 50, 25, and 12 nodes. To optimize the node weights, we utilize Adam, a stochastic gradient descent algorithm  \citep{2014arXiv1412.6980K}. For our non-linear activation function, we chose the rectified linear unit (ReLU), a reliable transformation used in most modern networks. Two key hyperparameters we consider in this work, which can be tweaked to improve performance, are $\alpha$ and the number of training iterations. $\alpha$ serves as the strength of the L2 regularization. If $\alpha$ is large, there is a high penalty for misclassification, leading to a greater likelihood of overfitting the training data. The number of training iterations is how often the weights of the MLP are updated. The larger this value, the more vulnerable the MLP is to overfitting.

To implement these models, we use Python's \texttt{scikit-learn} library \citep{scikit-learn}. This package provides excellent support for data preprocessing, feature selection, and hyperparameter tuning. 

\subsection{Data Preprocessing, Feature Selection, \& Hyperparameter Tuning}
\label{sec:preprocess}
Data preprocessing is a key aspect to consider when training ML models, as incorrectly formatted or inconsistent data can lead to significantly worse predictions. In particular, disagreement in feature scales (due to differences in units) can pose problems, since features with larger scales tend to be given additional weight. We address this problem by rescaling each training dataset feature distribution to a mean of 0 and a variance of 1, using \texttt{scikit-learn's Standard Scaler} module. This transformation is calculated using the following formula: $z = \frac{x-u}{s}$, where $z$  is the transformed value, $x$ is the input value, $u$ is the mean of the training samples, and $s$ is the standard deviation of the training samples. The transformation is then applied to the testing dataset, using the same training values for $u$ and $s$ to ensure that no testing data bias is introduced into our predictions.

Feature selection is another crucial facet to include, as utilizing features with little predictive capacity will result in poor performance. To determine the optimal features to select, we employ \texttt{scikit-learn’s SelectKBest} module, which calculates the analysis of variance (ANOVA) F-value for each feature in the training dataset. This univariate statistic provides an estimate for the separation of variances between two distributions (in this case flaring/non-flaring events). Narrow and widely spaced distributions will produce large F-values, while significantly overlapping distributions with large standard deviations will result in small F-values. The metric is mathematically defined in the following way \citep{bobra2015solar}: 
\begin{equation}
    F(i) = \frac{(\bar{x}_i^+-\bar{x}_i)^2+(\bar{x}_i^--\bar{x}_i)^2}{\frac{1}{n^+-1}\sum_{c=1}^{n^+}(\bar{x}_{c,i}^+-\bar{x}_i)^2+\frac{1}{n^--1}\sum_{c=1}^{n^-}(\bar{x}_{c,i}^--\bar{x}_i)^2}
\end{equation}
where, for a given feature $i$, $\bar{x}_i^+$ is the average value across flaring events, $\bar{x}_i^-$ is the average value across non-flaring events, $\bar{x}_i$ is the average value across all events, $n^+$ is the number of flaring events, and  $n^-$ is the number of non-flaring events. The $k$ features with the highest F-values are then kept, as they have the best-separated distributions, and thus, are beneficial to use when training our models. Determining the appropriate value for $k$ can be challenging, so several were tested: 1, 5, 10, 25, 50, and 120 (see Section \ref{sec:featuresandwindow}). We apply this selection methodology to every new training dataset, prior to undersampling (see Section \ref{sec:trainingwindows}). Ultimately, the main reason we settled on this feature selection approach over others is its proven success within SWAN-SF. Generally, F-values have been shown to provide reasonable insight into a feature's importance \citep{9679962}.

Lastly, to address hyperparameter tuning, a necessary step for maximizing predictive performance, we implement a grid search using \texttt{scikit-learn’s GridSearchCV} module. This enables us to exhaustively test combinations of hyperparameters and select those that result in the best-performing model. For each training dataset, a stratified group 5-fold cross-validation was applied. This ensures that within the training and testing folds, no data overlaps between ARs, and a similar number of flaring and non-flaring events are present. Using the generated folds, a model for every possible combination of hyperparameters shown in Table \ref{tab1} was tested. The model that produced the highest true skill statistic score (see Section \ref{sec:performancemetrics}) was then selected for application to the full training dataset. Once again, we apply this process for all new data fed to the model.

\begin{deluxetable*}{l|l}[htb]
\tablehead{\colhead{ML Classifier} & \colhead{\texttt{scikit-learn} Grid Search Hyperparameters}}
\startdata
Decision Tree & \texttt{criterion: ["gini", "entropy"]}\\
              & \texttt{class\_weight: ["balanced"]}\\
              & \texttt{max\_depth: [2, 3, 4, 5, 10, 20, 30, 40, 50, 100]}\\
              & \texttt{min\_samples\_leaf: [1, 10, 20, 30, 40, 50, 100]}\\
              & \texttt{min\_samples\_split: [2, 10, 20, 30, 40, 50, 100]}\\
\hline
Support Vector Machine & \texttt{kernel: ["rbf"]}\\
                       & \texttt{class\_weight: ["balanced"]}\\
                       & \texttt{C: [0.0001, 0.001, 0.01, 0.1, 1, 10, 100, 1000]}\\
                       & \texttt{gamma: [scale, 0.0001, 0.001, 0.01, 0.1, 1, 10, 100, 1000]}\\
\hline
Multilayer Perceptron & \texttt{hidden\_layer\_sizes: [(50, 25, 12)]}\\
                      & \texttt{solver: ["adam"]}\\
                      & \texttt{activation: ["relu"]}\\
                      & \texttt{learning\_rate: ["adaptive"]}\\
                      & \texttt{alpha: [0.0001, 0.001, 0.01, 0.1, 1, 10, 100]}\\
                      & \texttt{max\_iter: [5, 10, 20, 30, 40, 50, 100, 200]}
\enddata
\tablecomments{The SVM \texttt{scale} hyperparameter for \texttt{gamma} uses the inverse of the number of features times the variance of the feature vector. See the \texttt{scikit-learn} API (\href{https://scikit-learn.org/stable/modules/classes.html}{https://scikit-learn.org/stable/modules/classes.html}) for additional details on each parameter.}
\label{tab1}
\caption{A list of hyperparameters used in the DT, SVM, and MLP grid search.}
\end{deluxetable*} 

\subsection{Simulated Real-time Training Windows}
\label{sec:trainingwindows}
\begin{figure*}[htb]
    \centering
    \gridline{\fig{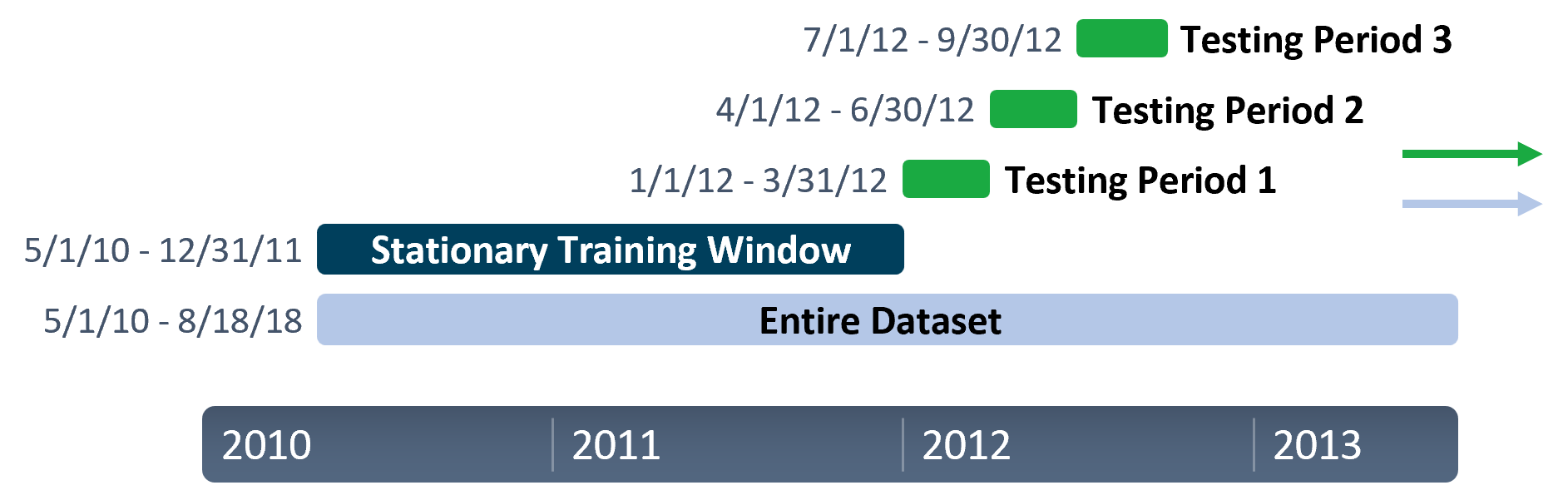}{.87\textwidth}{(a)}}
    \gridline{\fig{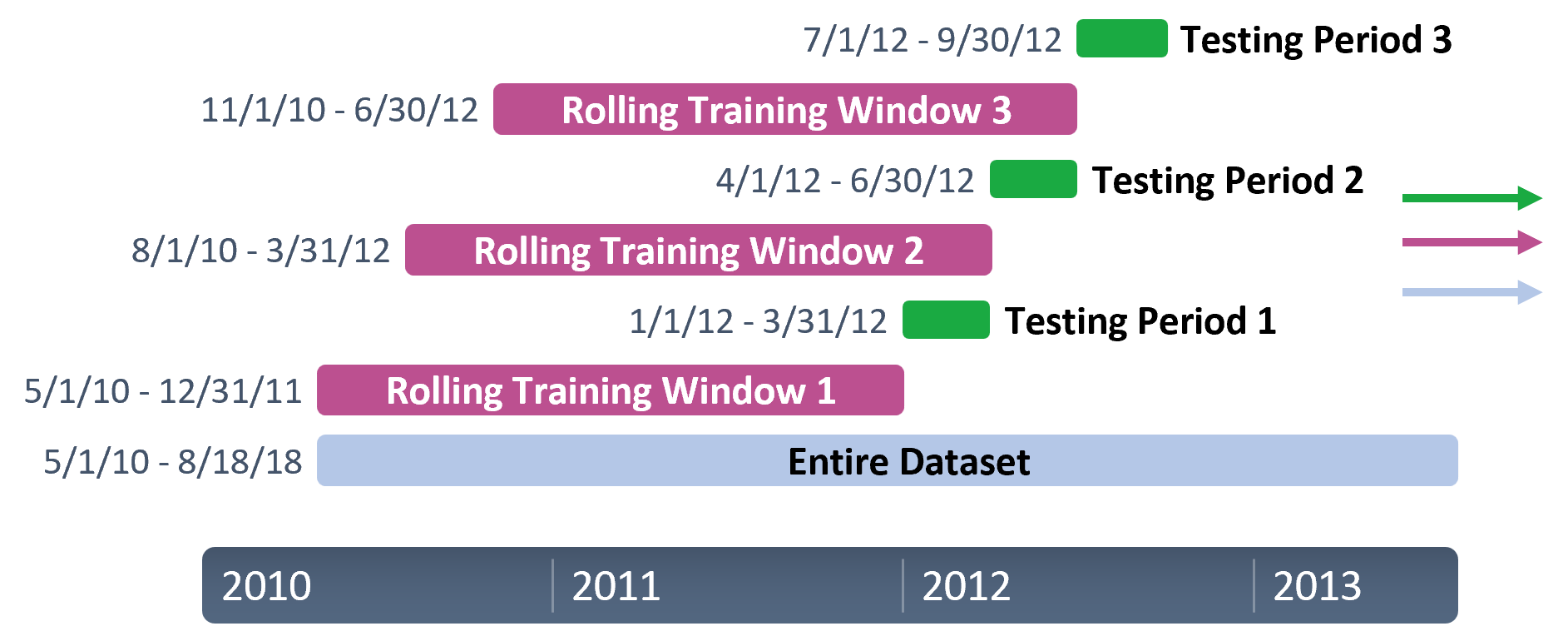}{.87\textwidth}{(b)}}
    \gridline{\fig{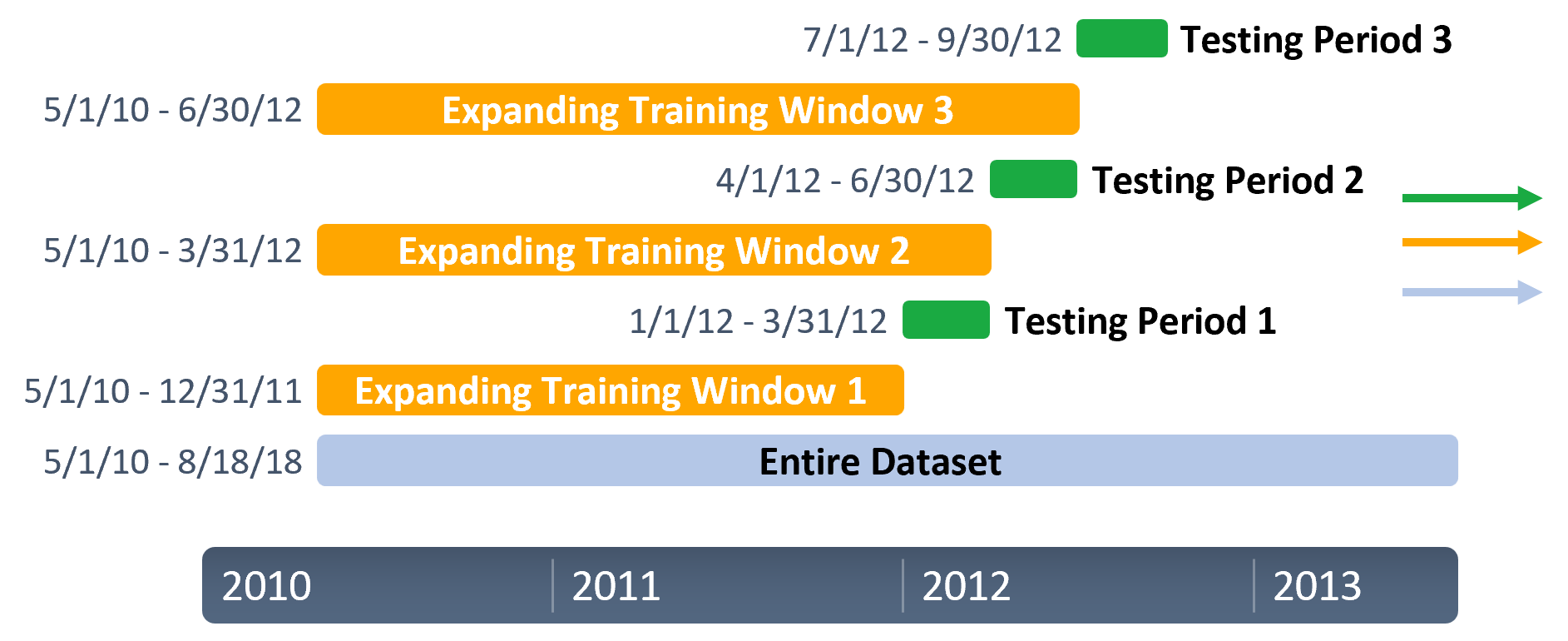}{.87\textwidth}{(c)}}
    \caption{(a) An example of the stationary training window methodology. Each model is trained on a set portion of the data at the beginning of the solar cycle (in this example it is the first 20 months). Model performance is then analyzed in consecutive 3-month blocks after training. (b) An example of the rolling training window methodology. Each model is trained similarly to the stationary window, however, the window now moves with the testing blocks. (c) A depiction of the expanding window methodology. Here, the models are trained using the entire available dataset, prior to the forecasting instance. The arrows in each figure emphasize the temporal continuation of the training windows, testing windows, and the dataset itself.}
    \label{fig:trainingwindows}
\end{figure*}
To explore the performance of a classifier in an operational setting, we designed a simulated real-time environment centered on training and testing ML models chronologically throughout Solar Cycle 24. Training data were produced using three different dynamic temporal windows: stationary, rolling, and expanding (see Figure \ref{fig:trainingwindows}). 
The stationary window paradigm generates forecasts using a single set of data that is available before the \emph{first} forecasting instance. This training data is always selected from the beginning of the solar cycle (May 2010 onward). The rolling window paradigm generates forecasts based on data from a constant time interval before the currently observed forecasting instance. This is similar to the stationary window, however, now, the window moves with the testing data. Lastly, models trained using the expanding window paradigm utilize all available data before the currently observed forecasting instance. 

The boundary conditions for a given window were defined to best emulate data acquired in real-time. For a given window lower boundary date, denoted as $X$, all data with time series start dates $\geq X$ were retained. For a given window upper boundary date, denoted as $Y$, non-flaring data whose 24-hour forecasting window ends $\leq Y$ were kept. Data instances with forecasting windows extending beyond this date were excluded, as operators would need to wait the full 24 hours to confirm an AR as non-flaring. In the case of flaring data, time series can instantly be labeled as a flaring event, once an M or X class flare occurs. Thus, data corresponding to flares that took place at times $\leq Y$ were kept, even if their 24-hour forecasting window had extended beyond the boundary. 

For the testing data, windows were generated in sequential 3-month blocks starting January 1st, 2012. The boundaries for the blocks were set to encompass all flaring data and any non-flaring time series whose 24-hour forecasting window end date fell within the 3-month block. Any testing blocks that did not contain flaring data were not considered in our analysis of true skill statistic and Heidke skill score in Sections \ref{sec:featuresandwindow} and \ref{sec:windowsizes}. This includes the period between April 2016 and March 2017, as well as any time after September 2017.

For each training dataset, an undersampling approach was applied to mitigate the effects of class imbalance. Within a given window, all flaring data were retained. However, non-flaring data were randomly sampled to match the number of flaring events, while preserving the original ratio of C-class to B-class to flare quiet events. For example, consider a training window consisting of 119 X-class, 974 M-class, 5,481 C-class, 5,184 B-class, and 51,160 flare quiet data. There are a total of 1,093 flaring events, which we want to retain, and 61,825 non-flaring events, which we want to trim down. By calculating the ratio between the number of flaring and non-flaring events (1,093/61,825) and multiplying it by the number of C-class, B-class, and flare quiet events, we can determine the required sample size from each class to preserve their original ratio while adding up to the desired 1,093 non-flaring events. Consequently, when applying this technique to the previous example, we get 97 C-class, 92 B-class, and 904 flare quiet events. Generally, this approach has proven to be successful when training and testing with SWAN-SF \citep{ahmadzadeh2021train}.

Lastly, to investigate how performance scales with data volume, a series of stationary and rolling window sizes (5, 8, 11, 14, 17, and 20 months) were tested. For the stationary windows, data were selected starting from May 2010 up until the added window size. For the rolling windows, data were selected between the testing window start date and extended into the past by the rolling window size. For each model, a total of three trials were run, each with different randomly undersampled non-flaring data to ensure robust results. Models with training data lacking flaring events were disregarded, and instead, the previously available trained model would be used. This was only pertinent to the rolling window.

\subsection{Performance Metrics}\
\label{sec:performancemetrics}
To evaluate the performance of each classifier, the true skill statistic (TSS) and Heidke skill score (HSS$_2$) were calculated for every 3-month testing block. These metrics are defined in the following way \citep{bobra2015solar}: 
\begin{equation}
    TSS = \frac{TP}{TP+FN} - \frac{FP}{FP+TN}
\end{equation}
\begin{equation}
    HSS_2 = \frac{2\times[(TP\times TN)-(FN\times FP)]}{(TP+FN)\times(FN+TN)+(TP+FP)\times(FP+TN)}
\end{equation}
where TP = true positives (the number of correctly predicted flaring events), TN = true negatives (the number of correctly predicted non-flaring events), FP = false positives (the number of non-flaring events predicted as flaring), and FN = false negatives (the number of flaring events predicted as non-flaring).
TSS ranges from -1 to +1, with a score of 0 reflecting a classifier that makes random or purely positive/negative forecasts, a score of -1 reflecting a classifier that is always wrong, and a score of +1 reflecting a perfect classifier \citep{ahmadzadeh2021train}. This metric is particularly advantageous as it is unbiased to the class imbalance problem prevalent within flare forecasting \citep{bobra2015solar}. When using TSS, one must keep in mind that two models with the same score do not necessarily produce an identical number of true positives and true negatives. This is because the metric is dependent on the balance of the true positive rate ($\frac{TP}{TP+FN}$) and the false positive rate ($\frac{FP}{FP+TN}$), which can be individually tweaked to achieve the same score \citep{ahmadzadeh2021train}.

Like TSS, HSS$_2$ ranges from -1 to +1 and provides an insight into a model's improvement over a random forecast. However, the minimum score is now dependent on the class imbalance ratio. As it reaches 1:1 (an equal number of flaring and non-flaring events), the lower boundary approaches -1 \citep{ahmadzadeh2021train}. A score of 0 is equivalent to a random classifier, a negative score is representative of a classifier that performs worse than random, and a score of +1 reflects a perfect classifier. We have selected this definition of the Heidke skill score over the original (HSS$_1$), highlighted in \citet{bobra2015solar}, as it tends to be less sensitive to the effects of class imbalance. Nevertheless, compared to TSS, both definitions are significantly more susceptible, with scores decreasing as the class imbalance ratio increases (see Figures 2 and 4 in \citealt{bobra2015solar} and \citealt{ahmadzadeh2021train} for an illustrative example). Overall, both metrics are widely used in the community, enabling others to make comparisons to this work, provided that they apply a similar methodology as shown here.

\subsection{Dependency On The Solar Cycle}
\label{sec:dependenceonsolarcycle}
Finally, to highlight performance dependencies on the solar cycle, we investigate the interplay between a model's false positive rate ($FPR = \frac{FP}{FP+TN}$) and the solar SXR background flux. We approach this by calculating the Spearman correlation between these parameters for each model utilizing 25 features. All window types, window sizes, classifiers, and trials were included. Following this, we back up our results by performing a study on how FPR is influenced by the largest flare class generated by an AR, and how the frequency of complex ARs changes throughout the solar cycle.

\section{Results \& Discussion} \label{sec:results}
In the following sections, we discuss the results of our work in detail. In Section \ref{sec:featuresandwindow}, we explore the effects of feature selection on model performance and highlight the magnetogram features most frequently chosen in our forecasts. In Section \ref{sec:windowsizes}, we compare model performance between the different window types and investigate how the temporal size of the stationary and rolling training windows affect our predictions. Finally, in Section \ref{sec:SXR_Cor}, we determine a correlation between the solar SXR background flux and the FPR. 

\subsection{Impacts Of Feature Selection}
\label{sec:featuresandwindow}
\begin{figure*}[htb]
    \centering
    \includegraphics[width=.95\textwidth]{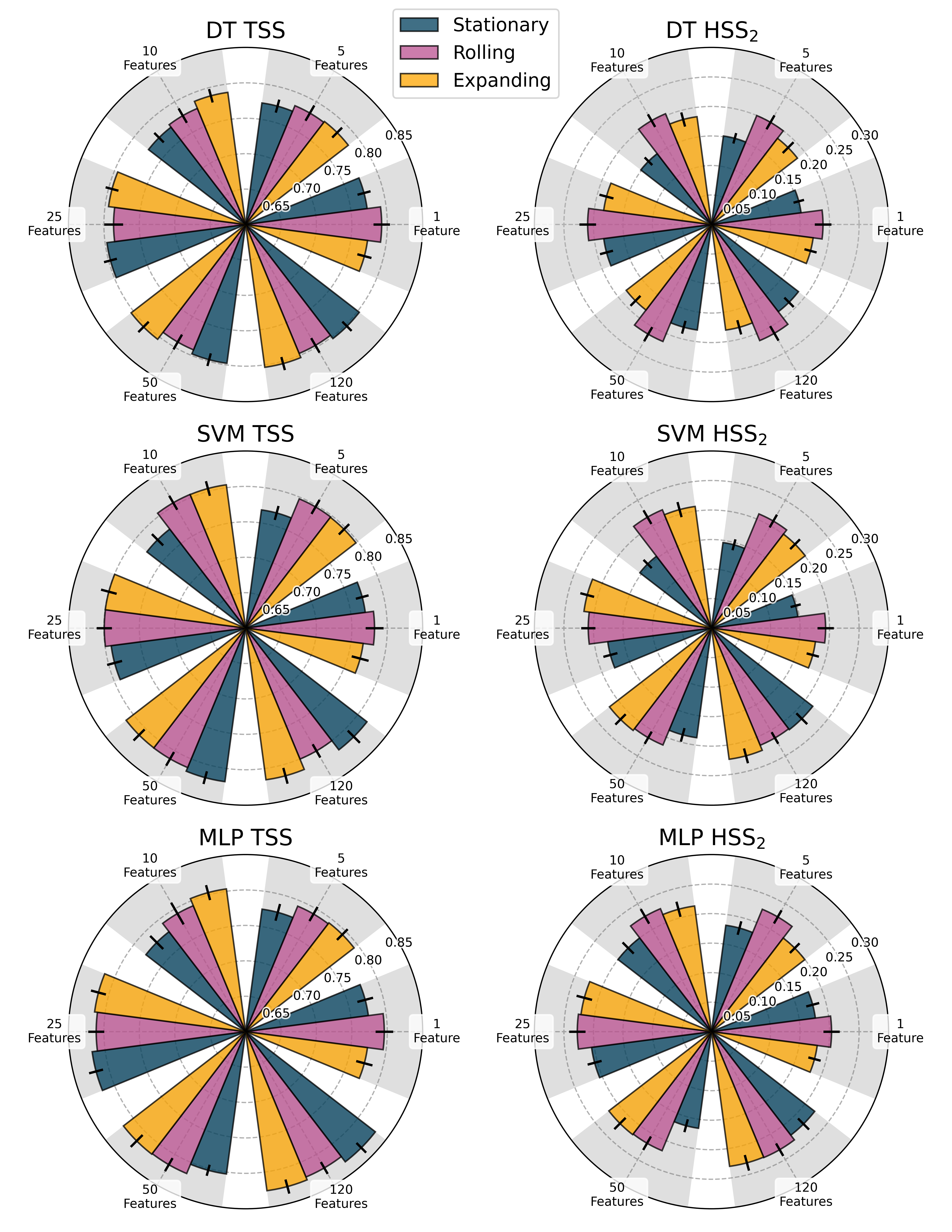}
    \caption{Average TSS and HSS$_2$ scores for the DT, SVM, and MLP classifiers with varying feature counts (1, 5, 10, 25, 50, 120) and window types. The error bars illustrate the standard error on the mean.\emph{ Note: These results were obtained using a stationary and rolling window of 20 months.}}
    \label{fig:featcomp}
\end{figure*}

The dimensionality of the feature space significantly influences the training time and complexity of a model. In an operational environment, unnecessary delays and complications must be avoided. Thus, we explore how the number of features selected for a given model affects performance, in hopes of establishing a baseline feature requirement for forecasts utilizing point-in-time magnetogram data. Figure \ref{fig:featcomp} summarizes our results. The columns highlight a particular skill score: TSS (left) and HSS$_2$ (right), while the rows correspond to our three tested classifiers: DT (top), SVM (middle), and MLP (bottom). The radar plots are divided into six sections, one for every feature count tested (1, 5, 10, 25, 50, and 120 features). Within each wedge, the radial extent of the bars denotes the skill score for a given feature set and window type (color), averaged across all available 3-month testing blocks (01/01/2012 - 03/31/2016, 04/01/2017 - 09/30/2017) associated with the 20-month stationary and rolling windows. This 20-month window was selected to remove any dependencies on data volume, which will be explored in Section \ref{sec:windowsizes}.

At first glance, we find that TSS and HSS$_2$ scores tend to increase as more features are included in a forecast. This is expected, given that a higher-dimensionality feature space offers additional means to distinguish between the flaring and non-flaring distributions. However, it is rather surprising that, for a given classifier and window type, skill scores improve on average by only 0.035 when jumping from 1 to 120 features. To check whether these improvements are statistically significant, we can compare the absolute difference between the two feature scores ($|\bar{X}_{120}-\bar{X}_1|$) and their combined standard errors ($\sqrt{\sigma_{\bar{X},120}^2+\sigma_{\bar{X},1}^2}$). When we do this, we find that 88.8\% of the improvements have a larger absolute difference than their combined errors, implying that they are statistically significant measurements at a 1$\sigma$ or 68\% confidence interval. If we extend this to 2$\sigma$, we discover that over 61\% of the improvements are statistically significant at a 95\% confidence interval. This suggests that, in general, our observations are meaningful and not simply due to the uncertainties in our data. However, the general similarity between scores still warrants further investigation.

\begin{figure*}[htb]
    \includegraphics[width=\textwidth]{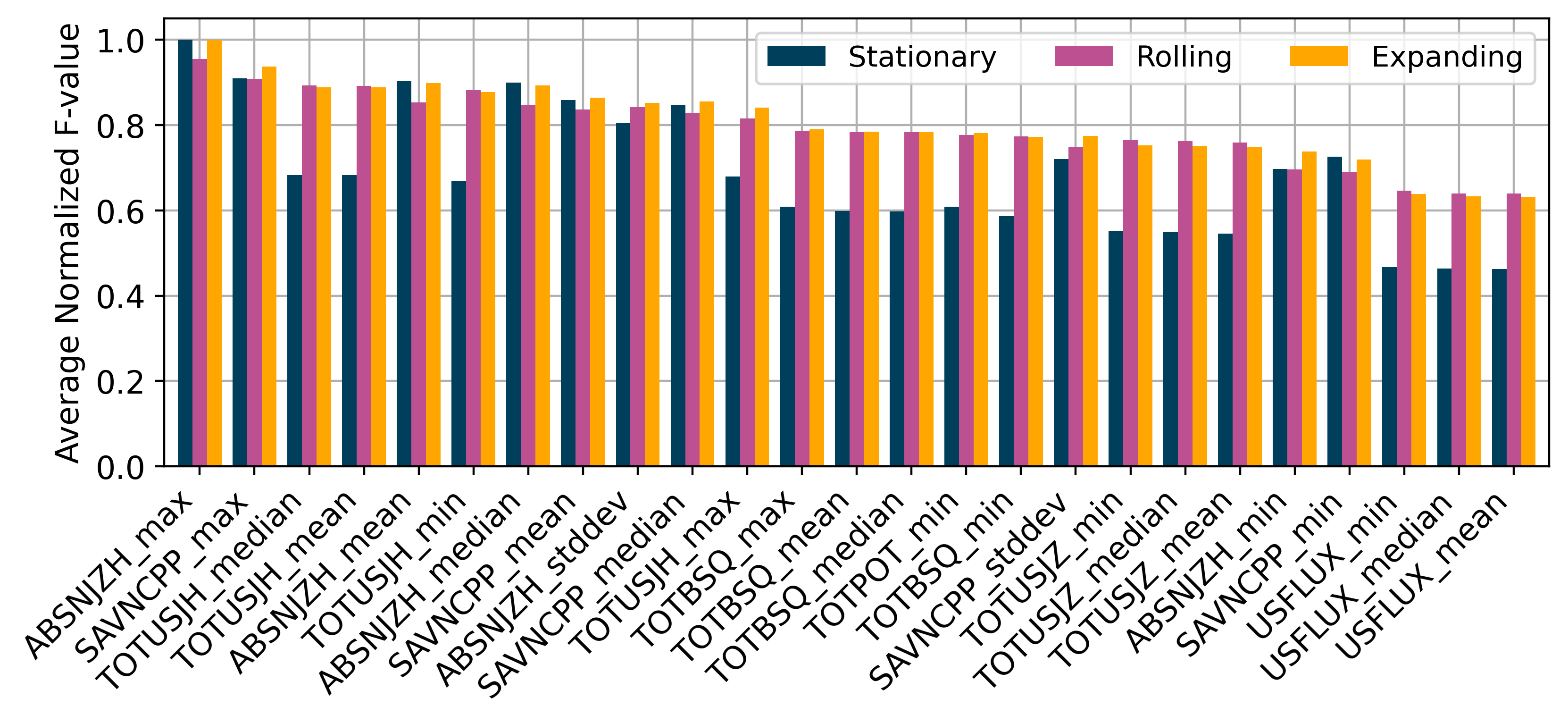}
    \caption{The average normalized F-value for the 25 highest-scoring features. This metric can be thought of as a proxy for selection frequency. Features with values close to 1 tend to be those with the highest scoring F-value (and thus more likely to be chosen) in a given training window.}
    \label{fig:selectionfreq}
\end{figure*}

To explore this topic in more detail, we examine the features that are typically chosen for a forecast, the flaring and non-flaring populations of those features, and the correlations between them. Figure \ref{fig:selectionfreq} depicts the average normalized F-value for the 25 highest-scoring features. A metric, which can be considered a proxy for the selection frequency of a particular parameter. To calculate this measure, we first determined the F-value of each feature across all stationary, rolling, and expanding training datasets. For a particular dataset, all F-values were normalized with respect to the highest achieved F-value. The scores for each feature were then averaged over all datasets and finally organized in decreasing order. Please note that even though we display the averages for individual window types, the order shown is solely based on the average across all training datasets. Since there is only 1 instance of the stationary window, and 19 instances (one for each new testing dataset) for both the rolling and expanding windows, the stationary window only provides a small contribution to this order. From the figure, it is evident that the summary statistics from only a few magnetogram parameters tend to be chosen for a given forecast: \textbf{ABSNJZH} (absolute values of the net current helicity), \textbf{SAVNCPP} (sum of the absolute value of net current polarity), \textbf{TOTUSJH} (total unsigned current helicity), \textbf{TOTBSQ} (total magnitude of the Lorentz force), \textbf{TOTPOT} (total photospheric magnetic free energy density), \textbf{TOTUSJZ} (total unsigned vertical current), and \textbf{USFLUX} (total unsigned flux). These features align well with those highlighted in other work within the field \citep{bobra2015solar, 9679962, zhang2022solar}.

\begin{figure*}[htb]
    \centering
    \includegraphics[width=\textwidth]{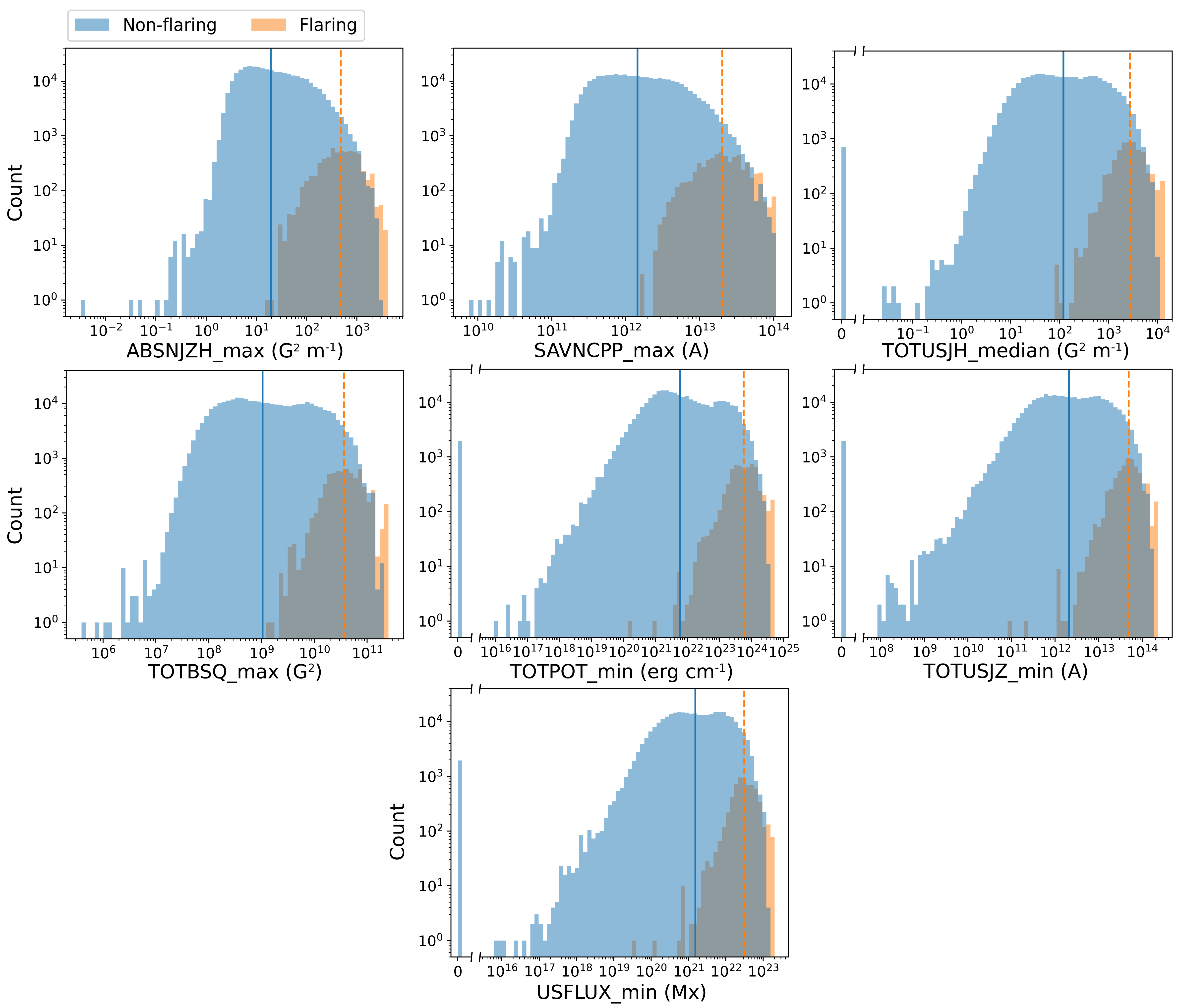}
    \caption{The flaring and non-flaring distributions for ABSNJZH\_max (the maximum absolute value of the net current helicity), SAVNCPP\_max (the maximum sum of the absolute value of net current polarity), TOTUSJH\_median (the median total unsigned current helicity), TOTBSQ\_max (the maximum total magnitude of the Lorentz force), TOTPOT\_min (the minimum total photospheric magnetic free energy density), TOTUSJZ\_min (the minimum total unsigned vertical current), and USFLUX\_min (the minimum total unsigned flux) over the entire SWAN-SF dataset. Distributions are plotted on a log-log scale. Bins containing zeros are plotted before the break in the x-axis. The solid blue line indicates the median of the non-flaring distribution. The dotted orange line indicates the median of the flaring distribution.}
    \label{fig:distributions}
\end{figure*}

For a more comprehensive look, we have plotted the flaring and non-flaring distributions of the highest-ranking statistics for these features (see Figure \ref{fig:distributions}). These figures reveal an overarching similarity between the distributions, with each of them having a right-skewed non-flaring and left-skewed flaring population. While significant overlaps exist, demonstrating the challenge behind flare forecasting, a separation between the medians of these populations can still be resolved. This provides enough distinction to make reasonable forecasts utilizing even a single feature (most frequently \textbf{ABSNJZH\_max}), which explains the relatively high TSS scores we have obtained.

\begin{figure*}[htb]
    \centering
    \includegraphics[width=\textwidth]{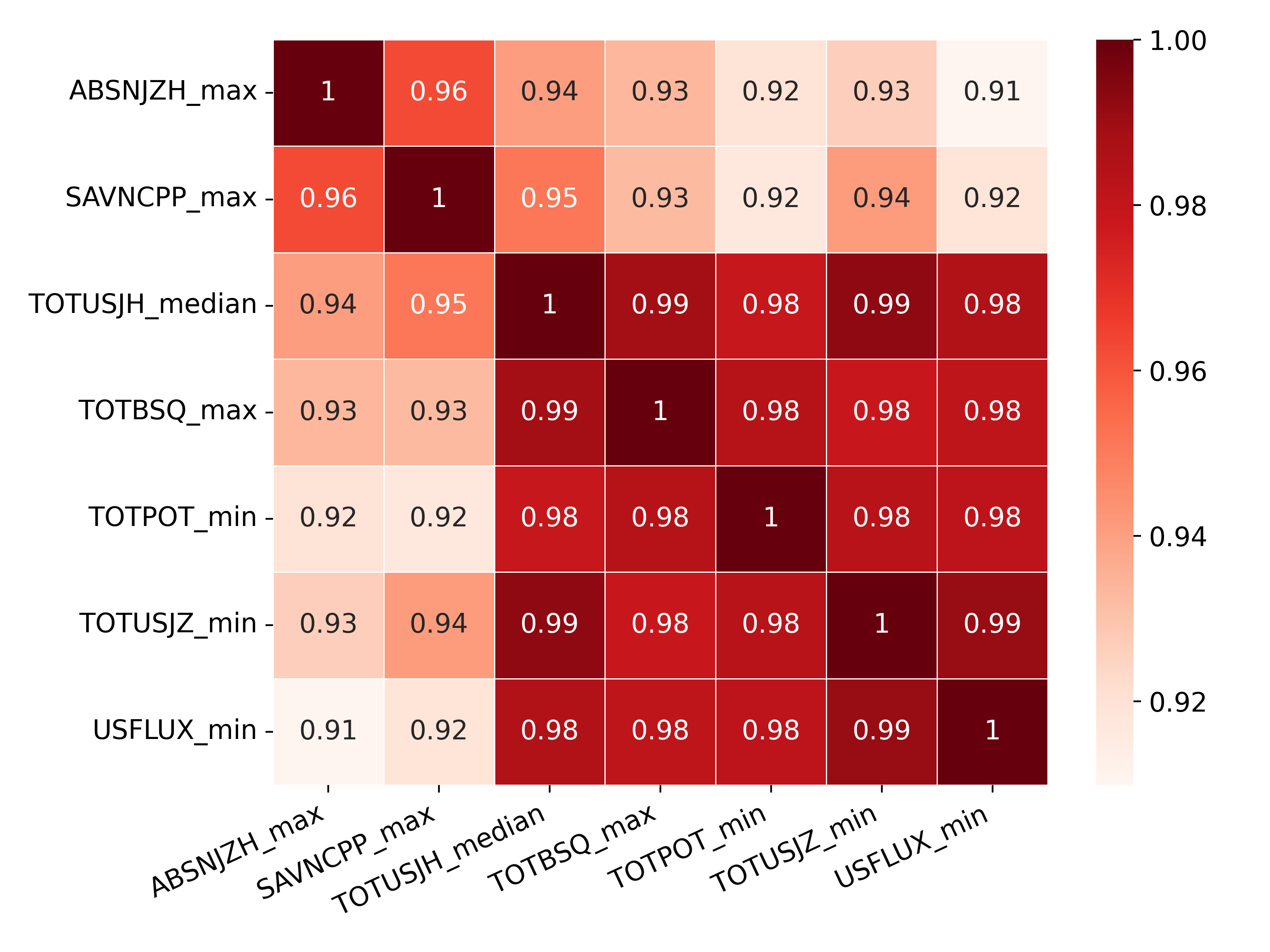}
    \caption{A Spearman correlation heatmap for the features selected in Figure \ref{fig:distributions}.}
    \label{fig:correlations}
\end{figure*}

Calculating the Spearman correlation between the 7 features in Figure \ref{fig:distributions}, we find that a strong positive correlation exists between all of them (see Figure \ref{fig:correlations}). Extending this analysis to all 25 features in Figure \ref{fig:selectionfreq}, it comes as no surprise that a similar trend is found, with all unique correlations being $\geq$0.72 and 80\% of them being $\geq$0.90. Though correlation does not necessarily mean that two features aren't complementary \citep{Guyon2003}, we believe that this could still be a plausible explanation for our results. Highly correlated features often provide similar information about the target class. Therefore, combining them will result in only minor improvements in the separability of the population. In our case, the 25 most important features are highly correlated, which may explain why the performance from 1 to 25 features is fairly comparable. Beyond 25, the additionally included features have diminishing F-values, making it significantly more difficult to separate between flaring and non-flaring events (at least in a 1-dimensional sense). This may explain the only marginal performance improvements found in this range.  

\begin{figure*}[htb!]
    \centering
    \includegraphics[width=\textwidth]{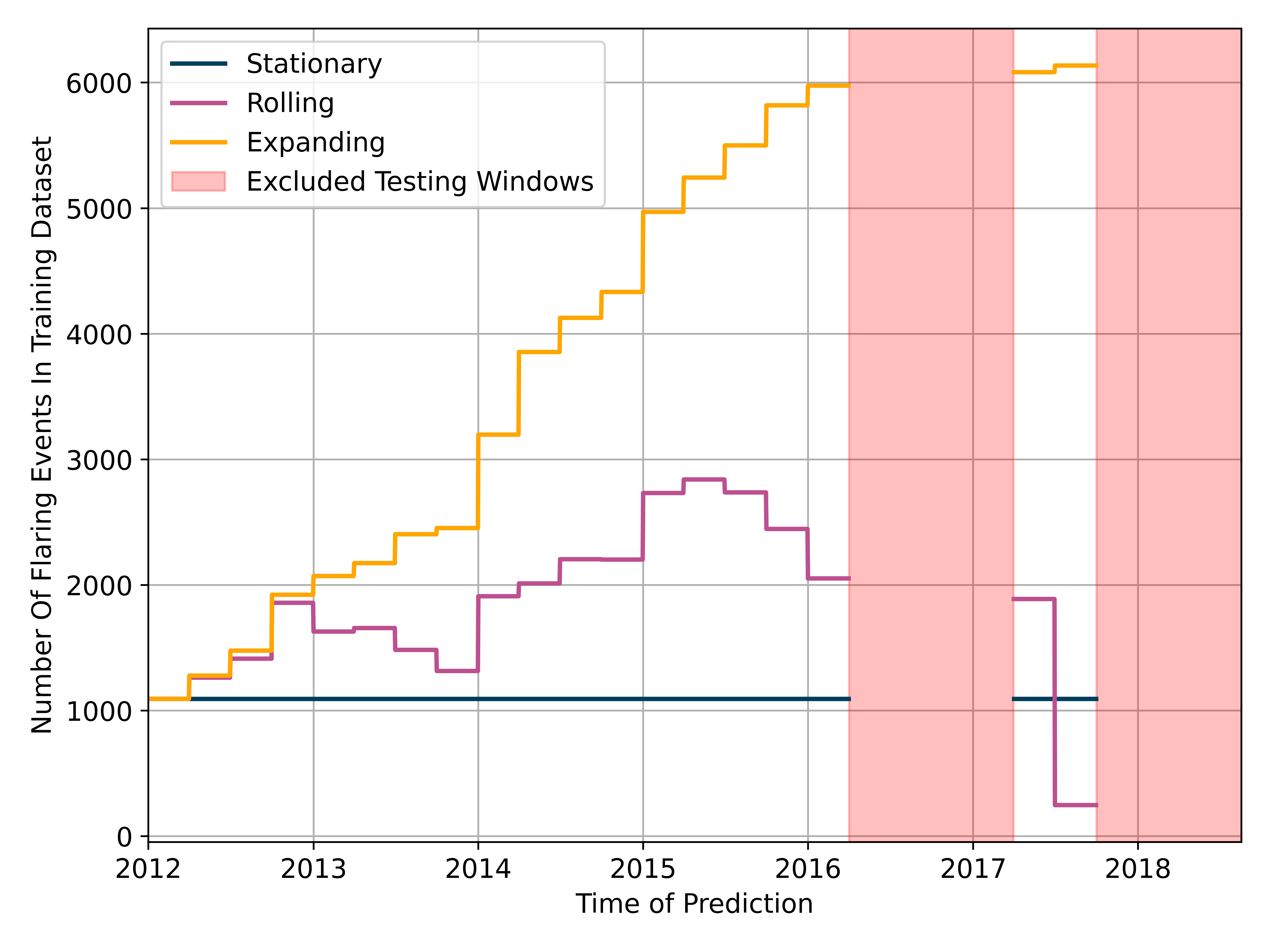}
    \caption{The number of flaring events as time progresses in the 20-month stationary, rolling, and expanding windows. The red regions illustrate periods where no testing windows exist, so no model was trained.}
    \label{fig:cumulativeflaringcount}
\end{figure*}

Shifting our focus to individual window types, we find that the rolling window consistently matches or outperforms the stationary and expanding windows, particularly when utilizing only 1 or 5 features. We speculate that this may be a consequence of the window's ability to capture the current flare occurrence rate. On a large scale, performance differences between the three window types are relatively minimal. This is unexpected given that, during the latter half of the solar cycle, the expanding window has access to significantly more flaring data than the other windows (see Figure \ref{fig:cumulativeflaringcount}). This suggests that, with a sufficient amount of data, a stationary classifier may be chosen over other window types. This not only saves time but dramatically reduces the difficulty of implementing an operational model. Of course, these results utilize a relatively large training window. Utilizing a smaller stationary or rolling window may not produce the same results. We explore this further in Section \ref{sec:windowsizes}.

\begin{figure*}[h!]
    \centering
    \includegraphics[width=.95\textwidth]{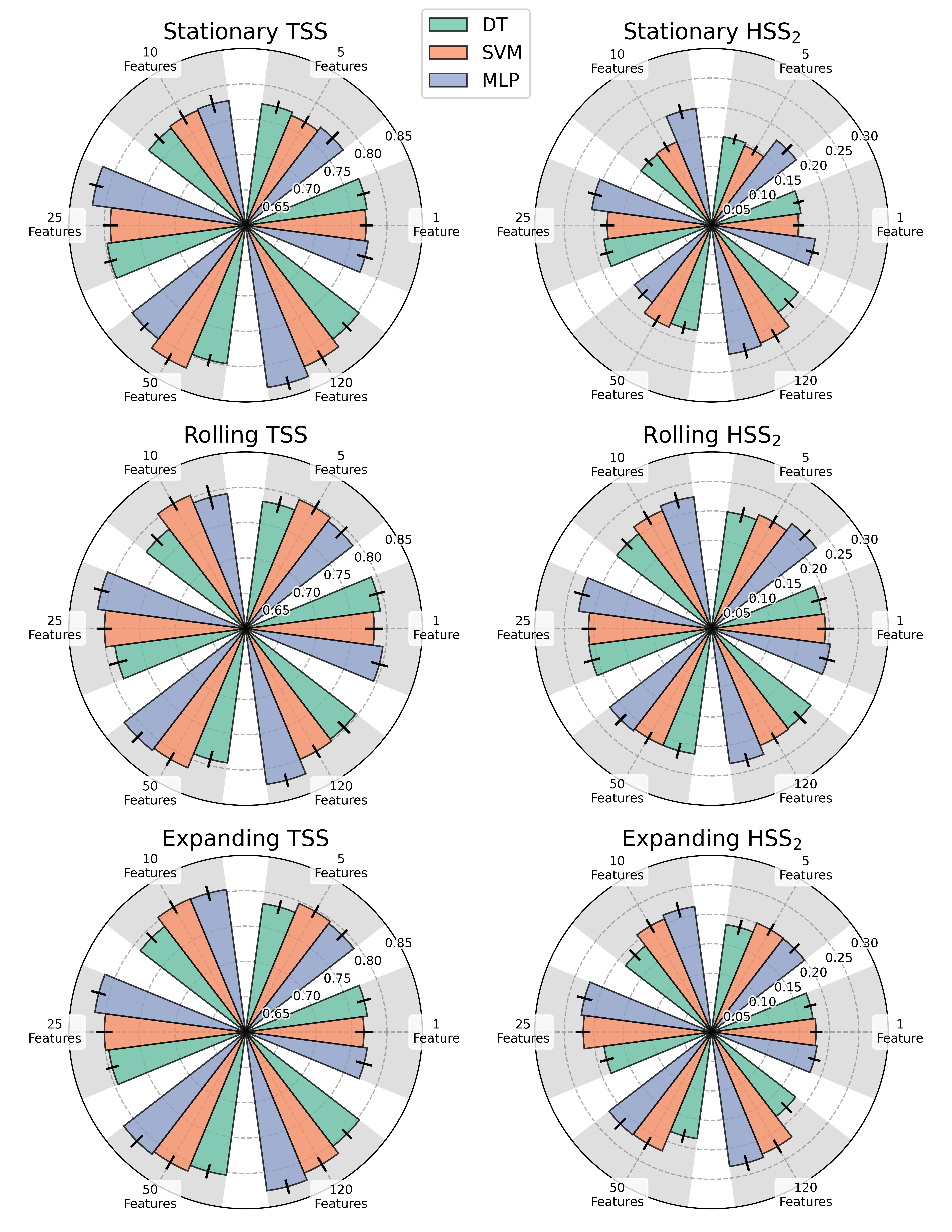}
    \caption{Average TSS and HSS$_2$ scores for the DT, SVM, and MLP classifiers with varying feature counts (1, 5, 10, 25, 50, 120) and window types. The error bars illustrate the standard error on the mean. This plot is similar to Figure \ref{fig:featcomp}, except that a comparison is now being made between classifiers instead of window type.\emph{ Note: These results were obtained using a stationary and rolling window of 20 months.}}
    \label{fig:featmodelcomp}
\end{figure*}

Finally, comparing results across the three tested classifiers, it becomes evident that MLPs yield the best TSS and HSS$_2$ scores, regardless of the number of features or window type selected (see Figure \ref{fig:featmodelcomp}). Given the algorithm's complexity, this is expected. However, the narrow difference between the skill scores of all three classifiers is rather surprising. This clearly suggests that easily interpretable models, such as DTs, may be a viable alternative to more complicated models when provided with a 20-month training window. We investigate this trend for different window sizes in Section \ref{sec:windowsizes}.

\subsection{Impacts Of Training Window Size}
\label{sec:windowsizes}

\begin{figure*}[htb]
    \centering
    \includegraphics[width=.95\textwidth]{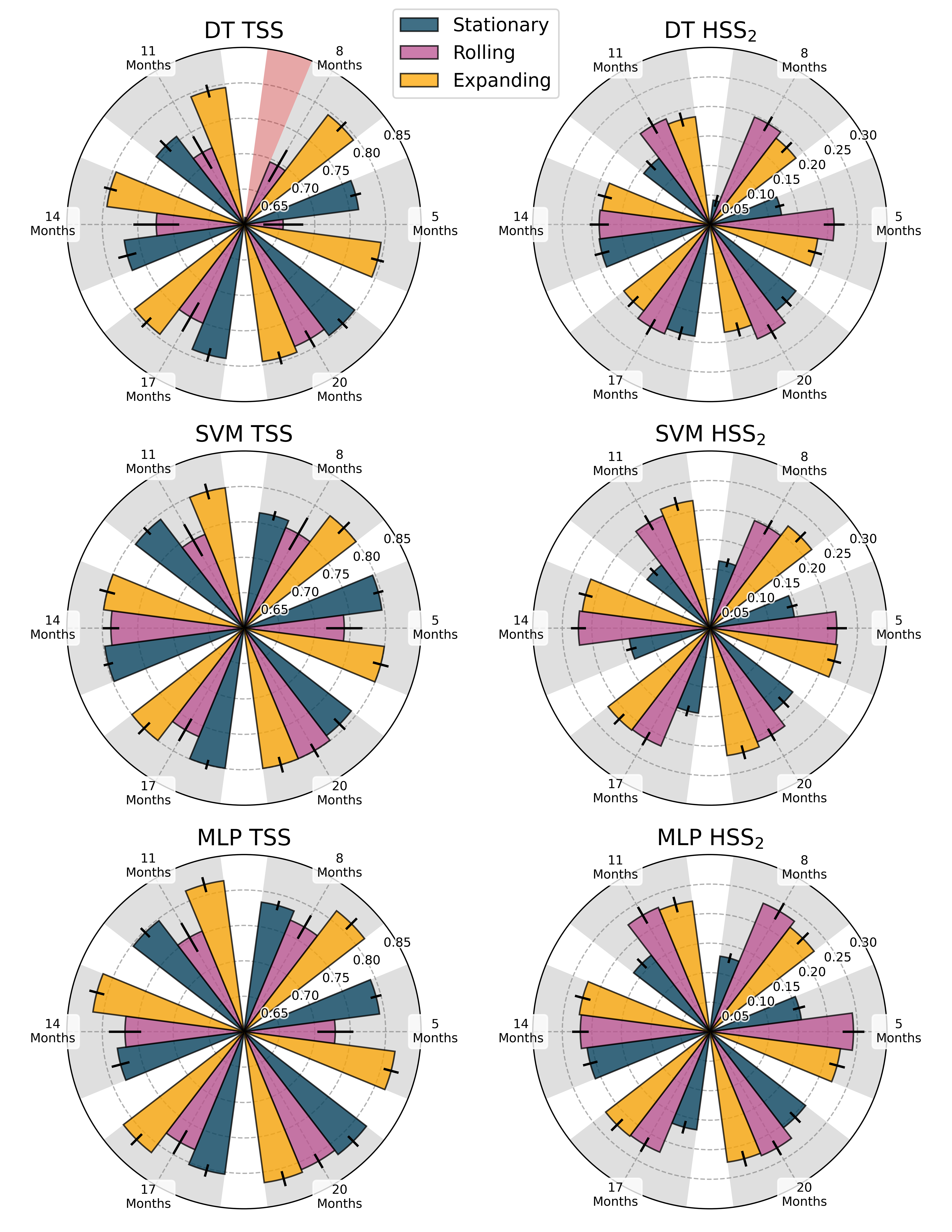}
    \caption{Average TSS and HSS$_2$ scores for the 25 feature DT, SVM, and MLP with varying stationary and rolling window sizes (5, 8, 11, 14, 17, 20 months). Naturally, the skill scores for the expanding window are the same across different window sizes. They are included for reference. The error bars illustrate the standard error on the mean.\emph{ Note: The 8-month stationary DT (the red wedge) has an average TSS score of 0.26 $\pm$ 0.05.}}
    \label{fig:time}
\end{figure*}

In addition to feature selection, data volumes are critical to producing effective flare forecasting models. Without the proper amount of training data, ML algorithms fail to capture an adequate decision boundary, which can significantly degrade performance. Since data volumes can vary during operational deployment, it is essential to explore how skill scores are affected by this aspect and how these restrictions interact with our custom training windows. Figure \ref{fig:time} summarizes our results. Once again, each column highlights a specific skill score, and each row a particular model. The radar plots are divided into six sections, one for every stationary and rolling window size tested (5, 8, 11, 14, 17, and 20 months). Within each wedge, the radial extent of the bars denotes the skill score for a given window size and type, averaged across all testing data (01/01/2012 - 03/31/2016, 04/01/2017 - 09/30/2017) associated with the 20-month stationary and rolling training windows. This ensures that our findings are comparable across window sizes. Since the expanding window utilizes all data prior to the forecasting instance, its scores are the same across all wedges.

First, examining the general skill score trends, we find that the effects of window size on TSS and HSS$_2$ are dependent on the classifier and window type selected. For certain combinations, such as the DT with rolling window, removing training data results in a steady decline in performance, as one might expect. However, we find that this is non-universal, with a majority of scores being completely uncorrelated with one another. For example, the stationary SVM TSS is larger for the 5-month window than the 20-month window, even though it has significantly fewer flares. This hints that there may be some underlying limitations to our dataset, which we suspect are imposed by our methodology. When calculating the summary statistics of each time series, we remove potentially significant knowledge related to the dynamics of an event. This gives us a lighter dataset, that is easier to work with but may not be as informative. It is apparent that our models are able to capture some important aspects needed to predict flares, as they achieve fairly high skill scores, but with additional data, this can only improve so much. Without more exhaustive features, which can be taken advantage of by our ML models, increasing data volumes will not provide enough new information about the flaring population to significantly affect performance. A potential solution to this problem is to train models utilizing the entire time series, which has been shown to improve skill scores in SWAN-SF \citep{timeseriespredSWAN}. However, it remains to be seen how, or even if, this would affect our data volume results. Lastly, a recent study has shown that magnetogram data alone does not provide significant improvements over human-based forecasting \citep{Leka_2019}. This hints that there may be some inherent simplicity to magnetogram data itself, which limits its predictive capability and contributes to the findings shown here.

Taking a deeper dive into our results, we find that the stationary window almost always produces better TSS, but noticeably worse HSS$_2$ scores, than the rolling window when the window size is less than 20-months. Since the stationary window covers the beginning of the solar cycle, where the number of flaring events is low, models will be biased toward capturing each flaring event in the training data. This is because missing a single flare has a large impact on the true positive rate ($\frac{TP}{TP+FN}$) and in turn the TSS score, which we are attempting to maximize when training the model. This leads to the stationary window producing fewer false negatives and more false positives while testing, which has less of an effect on TSS than HSS$_2$. 

\begin{figure*}[htb]
    \centering
    \includegraphics[width=.95\textwidth]{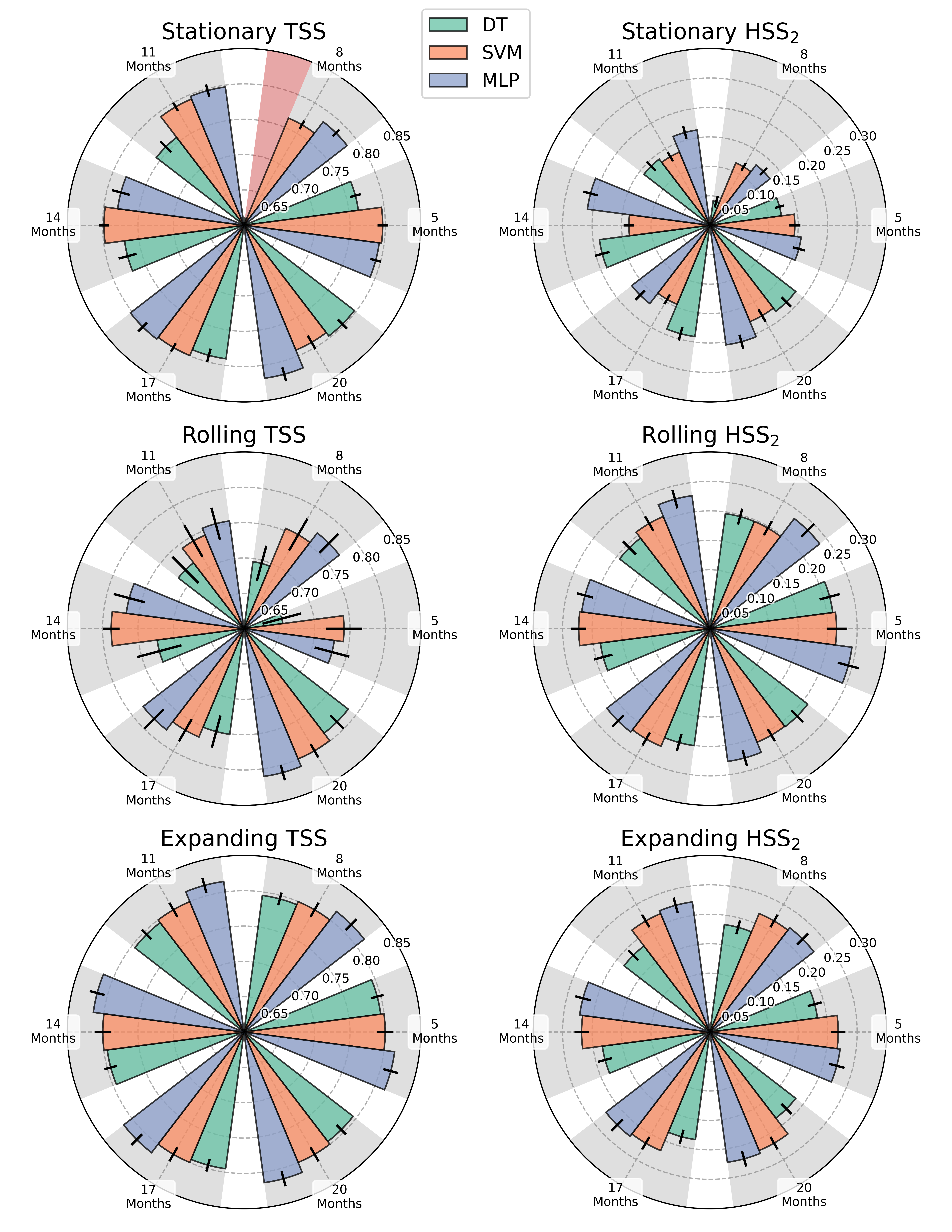}
    \caption{Average TSS and HSS$_2$ scores for the 25 feature DT, SVM, and MLP with varying stationary and rolling window sizes (5, 8, 11, 14, 17, 20 months). The error bars illustrate the standard error on the mean. This plot is similar to Figure \ref{fig:time}, except a comparison is now being made between classifiers instead of window type.\emph{ Note: The 8-month stationary DT (the red wedge) has an average TSS score of 0.26 $\pm$ 0.05.}}
    \label{fig:time2}
\end{figure*}

Lastly, when comparing performance across classifiers (see Figure \ref{fig:time2}), we again find that MLPs yield the best TSS and HSS$_2$ scores by only a small margin. SVMs and DTs, follow closely behind, even occasionally outperforming MLPs in select window types, sizes, and skill scores. This extends our conclusion made in \ref{sec:featuresandwindow}, that less complex models can be reliably used in place of more sophisticated algorithms, to varying data volumes. However, we find that when window sizes get too small, one must be cautious. The 8-month stationary DT window produces dreadful TSS and HSS$_2$ scores, which we suspect may be a consequence of the algorithm itself. It is well known that DTs tend to struggle with instabilities and under/overfitting, with small changes in their training data producing vastly different trees \citep{li2002instability}. These effects, compounded with the fact there is little training data, lead to a higher chance of producing a decision boundary that does not accurately separate the entire flaring population in the testing dataset.

\subsection{Dependency On The Solar Cycle}
\label{sec:SXR_Cor}
\begin{figure*}[htb]
    \centering
    \gridline{\fig{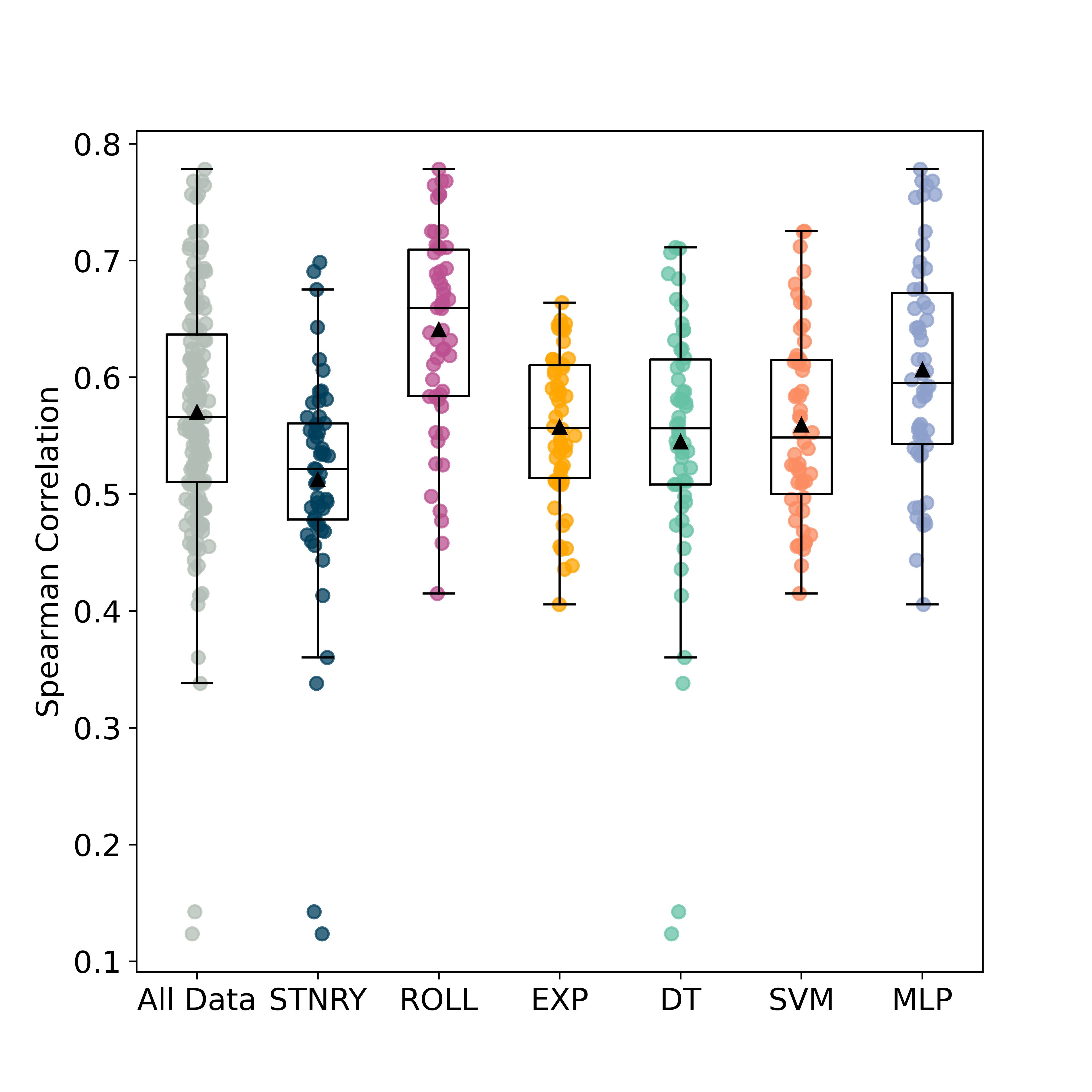}{.5\textwidth}{(a)}
            \fig{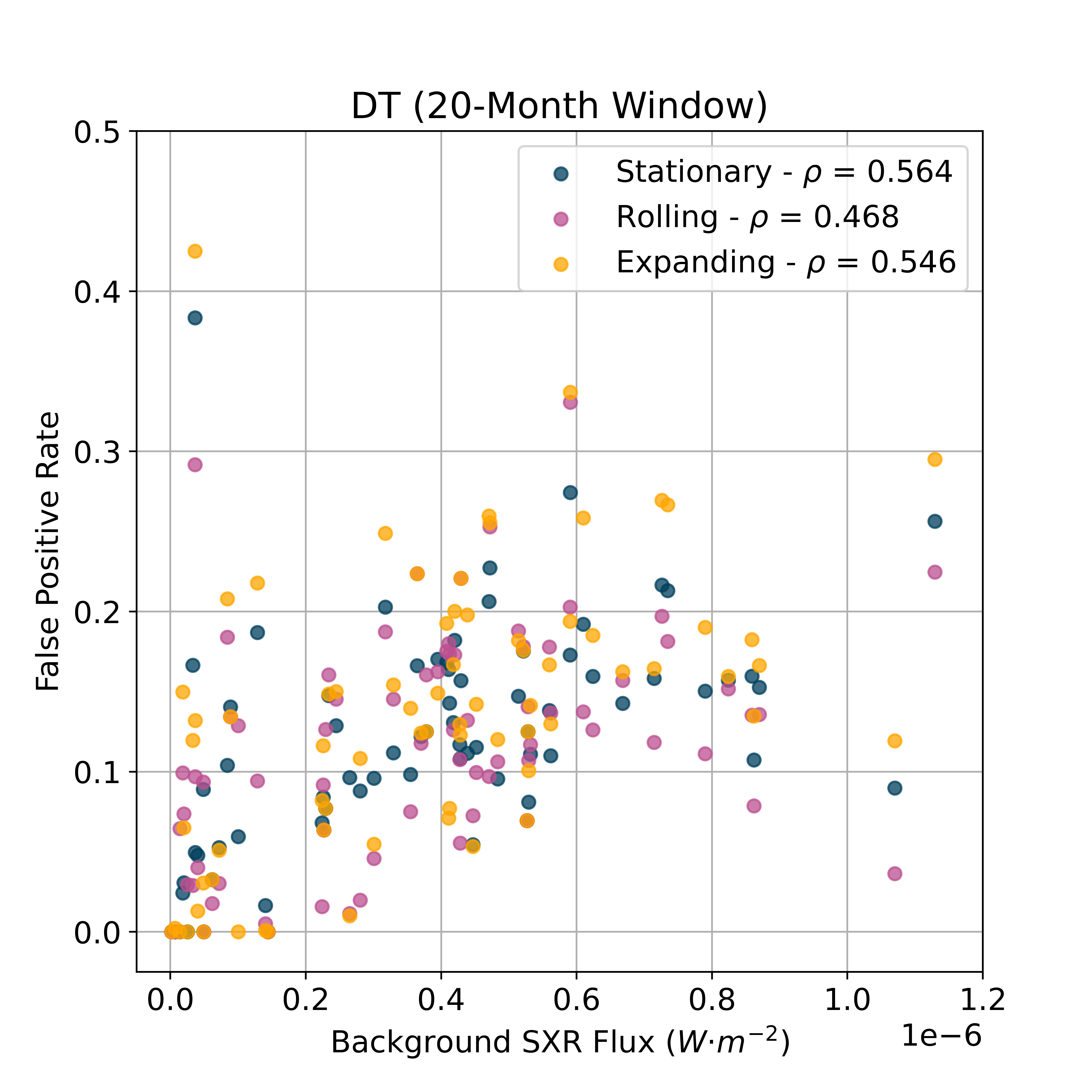}{.5\textwidth}{(b)}}
    \gridline{\fig{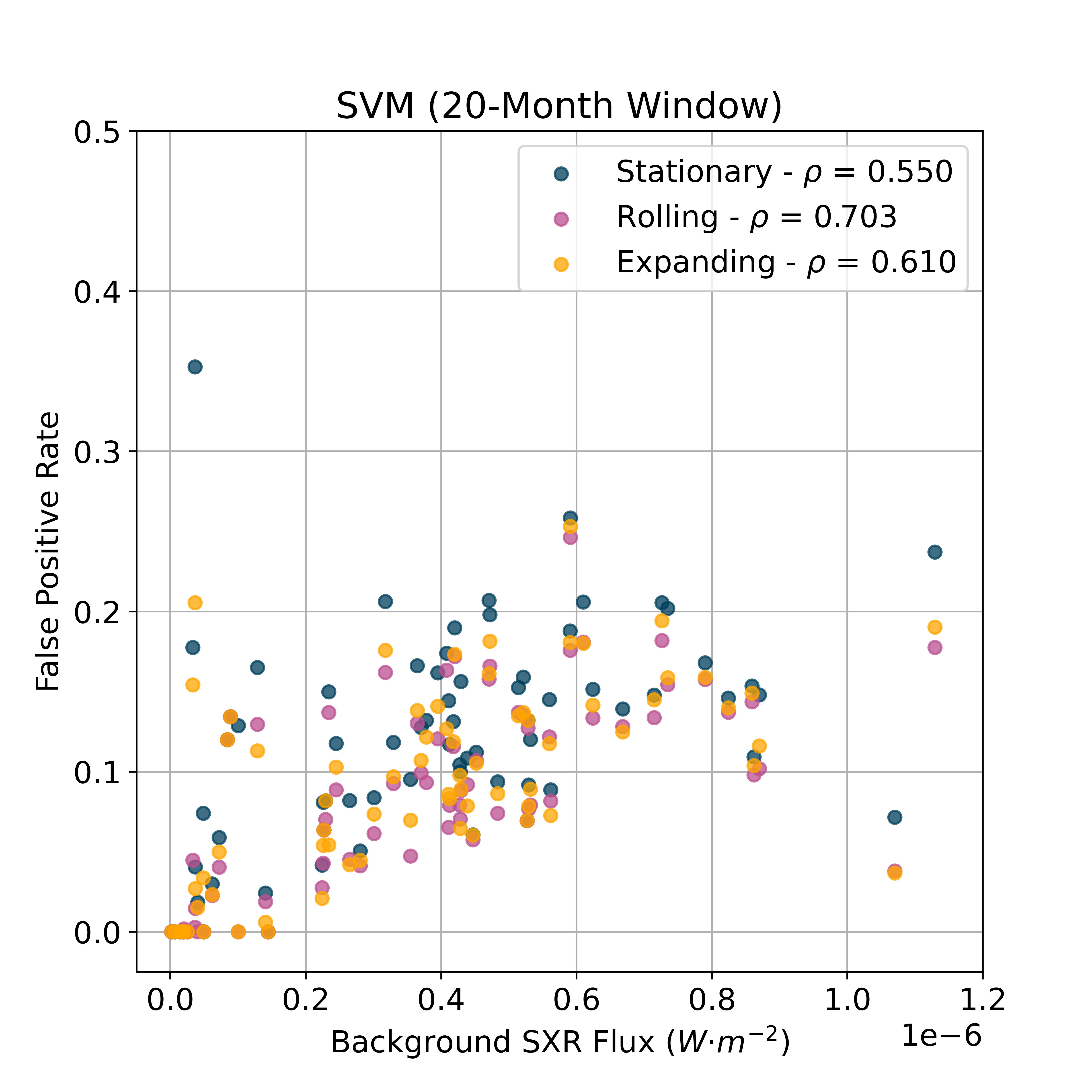}{.5\textwidth}{(c)}
            \fig{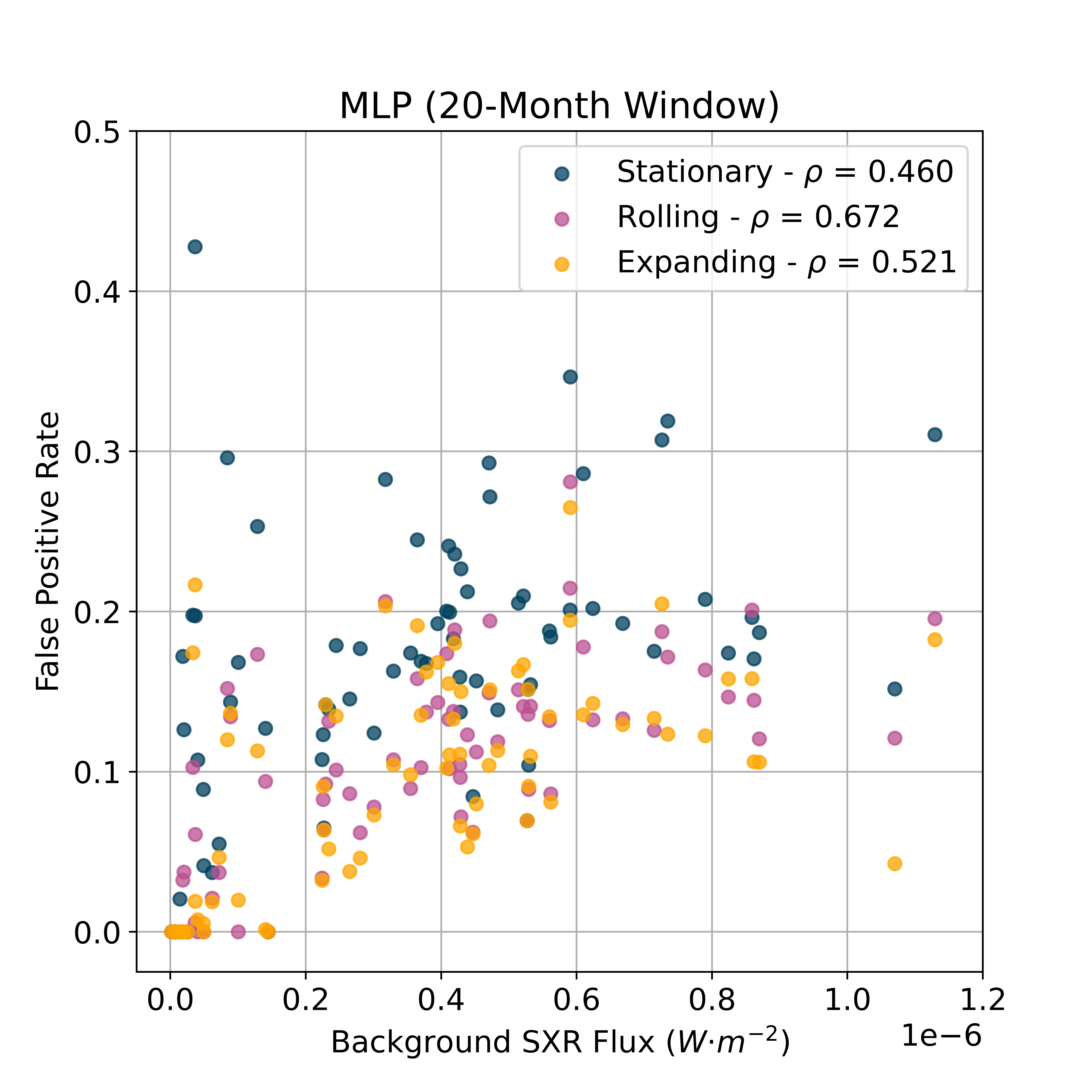}{.5\textwidth}{(d)}}
    \caption{(a) Boxplots of the Spearman correlation between the FPR and the background SXR flux. Plots are made for the entire collection of results, window types, and classifiers. Results are shown only for models that utilize 25 features. A swarm plot is overlaid to emphasize the distribution of data. The triangles indicate the mean of the distributions. (b, c, d) Scatter plots of the FPR versus background SXR flux for single trials of the 20-month DT, SVM, and MLP models. The correlation for each window type is given by $\rho$. The FPR and background SXR were binned monthly to reduce noise due to daily fluctuations.}
    \label{fig:boxplot}
\end{figure*}

Finally, to investigate the impact of the solar cycle on model performance, we explore the relationship between FPR and a proxy for the SXR background flux (the minimum daily GOES flux value). Figure \ref{fig:boxplot} summarizes our results. Here we find that a moderately strong positive correlation (mean of $\rho = 0.570$) exists for the conglomeration of our 25 feature trials (see the \emph{All Data} label in Figure \ref{fig:boxplot}). This indicates that as we reach solar maximum (when background levels are high), the FPR also increases. We find that models utilizing the rolling window or the MLP classifier tend to be more susceptible to this trend. Interestingly, no window type or classifier is impervious to this correlation.
\begin{deluxetable*}{c|c}[hbt]
\tablehead{\colhead{\shortstack{Largest Flare Produced By AR}} & \colhead{\shortstack{False Positive Rate Of ARs Within Flare Group}}}
\startdata
Flare Quiet / A & 0.010 $\pm$ 0.001 \\
\hline
B & 0.077 $\pm$ 0.004\\
\hline
C & 0.312 $\pm$ 0.007\\
\hline
M & 0.699 $\pm$ 0.008\\
\hline
X & 0.823 $\pm$ 0.008\\
\enddata
\label{tab2}
\caption{The FPR for ARs grouped by the strongest flare produced during their lifetime. Results are averaged across data from all models utilizing 25 features. The standard error on the mean is shown as well.}
\end{deluxetable*}

To explore this further, we then examine how the largest flare class generated by an AR influences its FPR. To accomplish this, we first divide our ARs into flaring groups dependent on the strongest event they produce within their lifetime (X, M, C, B, or A / flare quiet). Any point-in-time data associated with a particular AR is placed within the same group. We then recalculate the FPR for the data in each AR category and average our findings across all models utilizing 25 features. Table \ref{tab2} presents our results. It is evident that ARs producing M or X-class flares generally have elevated FPRs in comparison to ARs generating weaker flares. This is somewhat expected, given that B and C-class flares, as well as flare quiet periods, occur intermittently throughout flaring episodes, which can be challenging to detect. ARs may appear to have high-magnetic activity over the previous 12 hours (in comparison to a typical non-flaring event) but do not end up flaring. This, of course, leads to significantly more false positive predictions for these ARs.  

\begin{figure*}[htb]
    \centering
    \includegraphics[width=\textwidth]{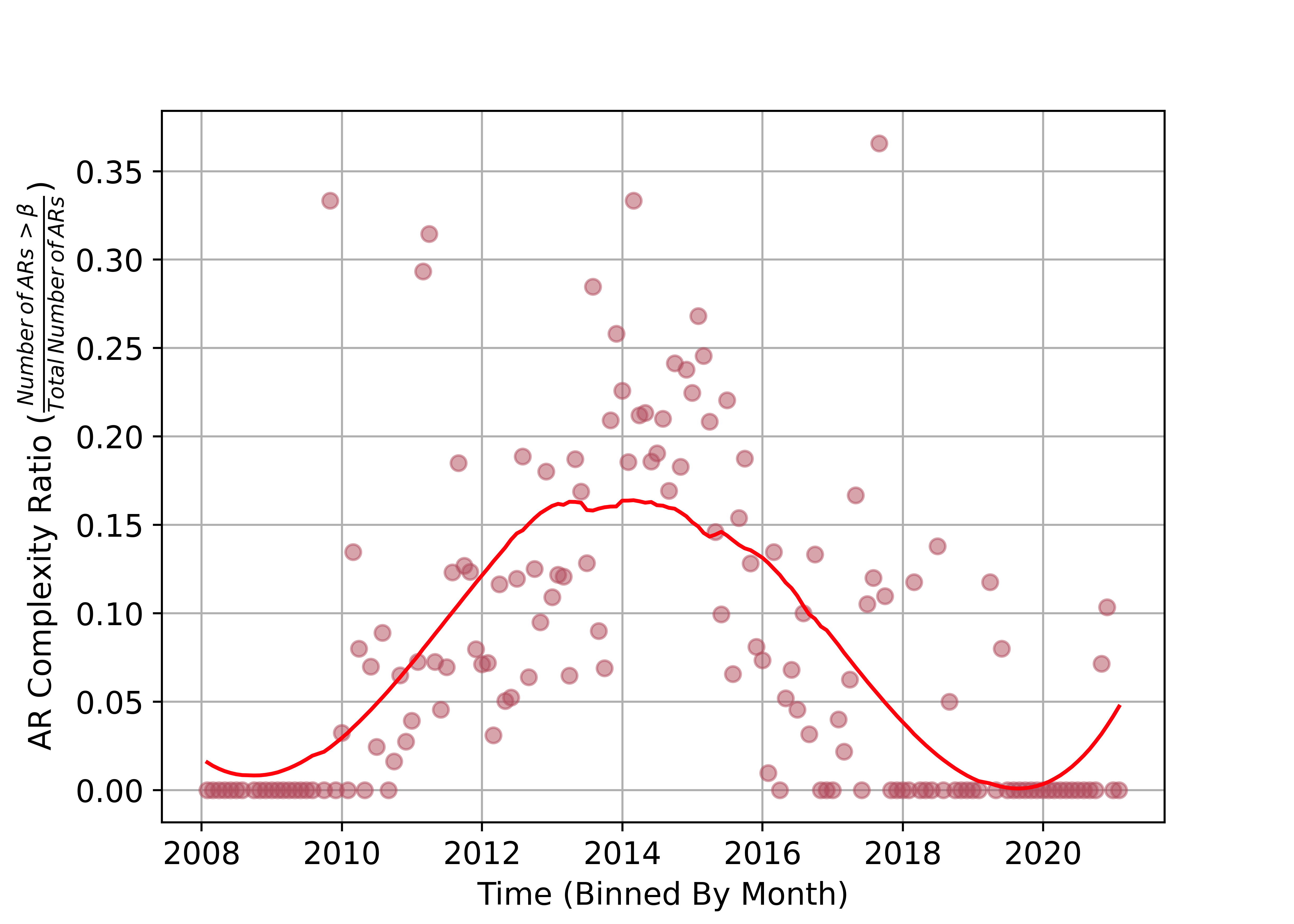}
    \caption{The AR complexity ratio versus time over Solar Cycle 24. AR complexity ratio is calculated by dividing the number of ARs with a Hale classification more complex than $\beta$  (this includes $\gamma$, $\beta-\gamma$, $\delta$, $\beta-\delta$, $\beta-\gamma-\delta$, and  $\gamma-\delta$) by the total number of ARs across each month. A Savitzky-Golay filter (solid red line) has been applied to illustrate the general trend of the data.}
    \label{fig:ARComplexity}
\end{figure*}

Building on this, we then consider the frequency of complex ARs (those more likely to produce M and X-class flares) throughout the solar cycle. In Figure \ref{fig:ARComplexity}, we plot the ratio of ARs with a Hale classification $>\beta$ (this includes $\gamma$, $\beta-\gamma$, $\delta$, $\beta-\delta$, $\beta-\gamma-\delta$, and  $\gamma-\delta$) to the total number of ARs, binned monthly for Solar Cycle 24. Here, we find that more complex ARs have a higher likelihood of existing during the peak of the solar cycle (near 2014) compared to the beginning or end. Tying this back to our findings from Table \ref{tab2}: if the probability of having a more complex AR is higher during the peak of the cycle, and ARs producing stronger flares tend to have larger FPRs, then there is reason to believe that the FPR will increase with background SXR flux. Of course, it is important to note that our forecasts are based solely on magnetic field parameters, with no direct relationship to the background SXR flux. Thus, we would like to emphasize that this result is merely a statistical observation rather than a causal relationship.

\section{Summary \& Conclusions}
\label{sec:conc}
In this study, we focused on producing a simulated real-time prediction environment, which can be used as a test bed to analyze how a variety of classifiers, features, data volumes, and the solar cycle impact operational performance. From this work, we have identified the following key results:
\begin{enumerate}
\item Across all window types, the most frequently chosen magnetogram features are \textbf{ABSNJZH} (absolute values of the net current helicity), \textbf{SAVNCPP} (sum of the absolute value of net current polarity), \textbf{TOTUSJH} (total unsigned current helicity), \textbf{TOTBSQ} (total magnitude of the Lorentz force), \textbf{TOTPOT} (total photospheric magnetic free energy density), \textbf{TOTUSJZ} (total unsigned vertical current), and \textbf{USFLUX} (total unsigned flux). This corresponds well with results from other papers within the field \citep{bobra2015solar, 9679962, zhang2022solar}.
\item The number of magnetogram features used to make a prediction does not have a significant effect on TSS or HSS$_2$ scores. Only a marginal increase in performance is observed as additional features are included in a forecast. We believe this may be an outcome of the highly correlated nature of our features.
\item When utilizing a 20-month stationary or rolling window, performance is generally comparable to the expanding window. Only a minor decrease in performance is observed for the stationary and rolling windows when their size is reduced. This suggests that, provided with a sufficient amount of data, a stationary classifier can be chosen over other window types, removing the need for retraining. We believe this to be a consequence of our methodology or potentially an inherent simplicity of the magnetogram data itself.
\item Simple and interpretable machine learning classifiers, such as decision trees, provide skill scores similar to those of more complex models.
\item A moderately strong positive Spearman correlation exists between a model's false positive rate and the background soft X-ray flux. We hypothesize that this is a consequence of highly complex active regions (those more likely to produce M and X-class flares) appearing more frequently during the peak of the solar cycle. From our analysis, we observed that these active regions tend to be accompanied by larger false positive rates.
\end{enumerate}
Overall, we can conclude that for operational forecasts utilizing point-in-time magnetogram data, the number of features, window size, window type, and classifier used have a minimal impact on performance, at least for those we tested. 

Regarding future studies, there are numerous paths we can explore. First, it may be valuable to investigate whether utilizing temporally dependent features, such as the time series derivative of a parameter, has any impact on the forecasting results shown here. These descriptive statistics could give a model better insight into how an active region is growing/decaying over time, which may lead to improved performance. However, recent work by \cite{nishizuka2017solar} found that these features are ineffective on time scales less than 24 hours. This indicates that we would likely need to extend our 12-hour observation window to benefit from them. A better alternative would be to train models directly on the time series data itself, rather than the point-in-time summary statistics. This could be accomplished through more complicated deep learning algorithms such as long short-term memory (LSTM) networks, which have been employed in other studies \citep{liu2019predicting,sun2022predicting}. Of course, with these models comes added training time and a need for increasingly powerful computational resources, which is not ideal for operational purposes. Lastly, a major drawback of the current SWAN-SF iteration is its focus on 24-hour forecasting. With the rapidly approaching NASA Artemis missions, it will be critical that we have the capability of predicting flaring events even farther in advance (hopefully up to 72 hours). While not addressed in this paper, it may be worthwhile in future studies to modify the current dataset labels for several extended forecasting windows (36, 48, 72 hours). This may reveal hidden intricacies between our various training methodologies, not found in this work.

\section{Acknowledgements}
We would like to thank the anonymous reviewer for their excellent feedback, which greatly improved the clarity of the text. This work was supported by NASA FINESST grant 80NSSC23K1639, NASA LWS grant 80NSSC22K0272, and NSF FDSS grant 1936361.

\bibliography{sample631}{}

\begin{thebibliography}{}
\expandafter\ifx\csname natexlab\endcsname\relax\def\natexlab#1{#1}\fi
\providecommand{\url}[1]{\href{#1}{#1}}
\providecommand{\dodoi}[1]{doi:~\href{http://doi.org/#1}{\nolinkurl{#1}}}
\providecommand{\doeprint}[1]{\href{http://ascl.net/#1}{\nolinkurl{http://ascl.net/#1}}}
\providecommand{\doarXiv}[1]{\href{https://arxiv.org/abs/#1}{\nolinkurl{https://arxiv.org/abs/#1}}}

\bibitem[{Ahmadzadeh {et~al.}(2021)Ahmadzadeh, Aydin, Georgoulis, Kempton, Mahajan, \& Angryk}]{ahmadzadeh2021train}
Ahmadzadeh, A., Aydin, B., Georgoulis, M.~K., {et~al.} 2021, The Astrophysical Journal Supplement Series, 254, 23

\bibitem[{Ali {et~al.}(2024)Ali, Sadykov, Kosovichev, Kitiashvili, Oria, Nita, Illarionov, O’Keefe, Francis, Chong, {et~al.}}]{ali2024predicting}
Ali, A., Sadykov, V., Kosovichev, A., {et~al.} 2024, The Astrophysical Journal Supplement Series, 270, 15

\bibitem[{Angryk {et~al.}(2020{\natexlab{a}})Angryk, Martens, Aydin, Kempton, Mahajan, Basodi, Ahmadzadeh, Cai, Filali~Boubrahimi, Hamdi, Schuh, \& Georgoulis}]{DV_2020}
Angryk, R., Martens, P., Aydin, B., {et~al.} 2020{\natexlab{a}}, {SWAN-SF}, V1,  Harvard Dataverse, \dodoi{10.7910/DVN/EBCFKM}

\bibitem[{Angryk {et~al.}(2020{\natexlab{b}})Angryk, Martens, Aydin, Kempton, Mahajan, Basodi, Ahmadzadeh, Cai, Filali~Boubrahimi, Hamdi, {et~al.}}]{angryk2020multivariate}
Angryk, R.~A., Martens, P.~C., Aydin, B., {et~al.} 2020{\natexlab{b}}, Scientific data, 7, 227

\bibitem[{Bobra \& Couvidat(2015)}]{bobra2015solar}
Bobra, M.~G., \& Couvidat, S. 2015, The Astrophysical Journal, 798, 135

\bibitem[{Camporeale(2019)}]{challengesML}
Camporeale, E. 2019, Space Weather, 17, \dodoi{10.1029/2018SW002061}

\bibitem[{{Crown}(2012)}]{crown2012noaaforecast}
{Crown}, M.~D. 2012, Space Weather, 10, S06006, \dodoi{10.1029/2011SW000760}

\bibitem[{{Deshmukh} {et~al.}(2023){Deshmukh}, {Baskar}, {Berger}, {Bradley}, \& {Meiss}}]{2023A&A...674A.159D}
{Deshmukh}, V., {Baskar}, S., {Berger}, T.~E., {Bradley}, E., \& {Meiss}, J.~D. 2023, \aap, 674, A159, \dodoi{10.1051/0004-6361/202245742}

\bibitem[{Florios {et~al.}(2018)Florios, Kontogiannis, Park, Guerra, Benvenuto, Bloomfield, \& Georgoulis}]{florios2018forecasting}
Florios, K., Kontogiannis, I., Park, S.-H., {et~al.} 2018, Solar Physics, 293, 28

\bibitem[{Gardner \& Dorling(1998)}]{GARDNER19982627}
Gardner, M., \& Dorling, S. 1998, Atmospheric Environment, 32, 2627, \dodoi{https://doi.org/10.1016/S1352-2310(97)00447-0}

\bibitem[{Guyon \& Elisseeff(2003)}]{Guyon2003}
Guyon, I., \& Elisseeff, A. 2003, J. Mach. Learn. Res., 3, 1157–1182

\bibitem[{Hudson(2021)}]{hudson2021carrington}
Hudson, H.~S. 2021, Annual Review of Astronomy and Astrophysics, 59, 445

\bibitem[{Ji {et~al.}(2020)Ji, Aydin, Georgoulis, \& Angryk}]{timeseriespredSWAN}
Ji, A., Aydin, B., Georgoulis, M.~K., \& Angryk, R. 2020, in 2020 IEEE International Conference on Big Data (Big Data), 4218--4225, \dodoi{10.1109/BigData50022.2020.9377906}

\bibitem[{{Kingma} \& {Ba}(2014)}]{2014arXiv1412.6980K}
{Kingma}, D.~P., \& {Ba}, J. 2014, arXiv e-prints, arXiv:1412.6980, \dodoi{10.48550/arXiv.1412.6980}

\bibitem[{Kingsford \& Salzberg(2008)}]{kingsford2008decision}
Kingsford, C., \& Salzberg, S.~L. 2008, Nature biotechnology, 26, 1011

\bibitem[{Kotsiantis(2013)}]{kotsiantis2013decision}
Kotsiantis, S.~B. 2013, Artificial Intelligence Review, 39, 261

\bibitem[{Leka {et~al.}(2019{\natexlab{a}})Leka, Park, Kusano, Andries, Barnes, Bingham, Bloomfield, McCloskey, Delouille, Falconer, {et~al.}}]{leka2019comparison}
Leka, K., Park, S.-H., Kusano, K., {et~al.} 2019{\natexlab{a}}, The Astrophysical Journal Supplement Series, 243, 36

\bibitem[{Leka {et~al.}(2019{\natexlab{b}})Leka, Park, Kusano, Andries, Barnes, Bingham, Bloomfield, McCloskey, Delouille, Falconer, Gallagher, Georgoulis, Kubo, Lee, Lee, Lobzin, Mun, Murray, Nageem, Qahwaji, Sharpe, Steenburgh, Steward, \& Terkildsen}]{Leka_2019}
Leka, K.~D., Park, S.-H., Kusano, K., {et~al.} 2019{\natexlab{b}}, The Astrophysical Journal, 881, 101, \dodoi{10.3847/1538-4357/ab2e11}

\bibitem[{Li {et~al.}(2007)Li, Wang, He, Cui, \& Du}]{li2007support}
Li, R., Wang, H.-N., He, H., Cui, Y.-M., \& Du, Z.-L. 2007, Chinese Journal of Astronomy and Astrophysics, 7, 441

\bibitem[{Li \& Belford(2002)}]{li2002instability}
Li, R.-H., \& Belford, G.~G. 2002, in Proceedings of the eighth ACM SIGKDD international conference on Knowledge discovery and data mining, 570--575

\bibitem[{Liu {et~al.}(2019)Liu, Liu, Wang, \& Wang}]{liu2019predicting}
Liu, H., Liu, C., Wang, J.~T., \& Wang, H. 2019, The Astrophysical Journal, 877, 121

\bibitem[{Marroquin {et~al.}(2023)Marroquin, Sadykov, Kosovichev, Kitiashvili, Oria, Nita, Illarionov, O’Keefe, Francis, Chong, Kosovich, \& Ali}]{Marroquin_2023}
Marroquin, R.~D., Sadykov, V., Kosovichev, A., {et~al.} 2023, The Astrophysical Journal, 952, 97, \dodoi{10.3847/1538-4357/acdb65}

\bibitem[{Natras {et~al.}(2019)Natras, Horozovic, \& Mulic}]{natras2019strong}
Natras, R., Horozovic, D., \& Mulic, M. 2019, SN Applied Sciences, 1, 1

\bibitem[{{Nishizuka} {et~al.}(2018){Nishizuka}, {Sugiura}, {Kubo}, {Den}, \& {Ishii}}]{2018ApJ...858..113N}
{Nishizuka}, N., {Sugiura}, K., {Kubo}, Y., {Den}, M., \& {Ishii}, M. 2018, \apj, 858, 113, \dodoi{10.3847/1538-4357/aab9a7}

\bibitem[{Nishizuka {et~al.}(2017)Nishizuka, Sugiura, Kubo, Den, Watari, \& Ishii}]{nishizuka2017solar}
Nishizuka, N., Sugiura, K., Kubo, Y., {et~al.} 2017, The Astrophysical Journal, 835, 156

\bibitem[{Pedregosa {et~al.}(2011)Pedregosa, Varoquaux, Gramfort, Michel, Thirion, Grisel, Blondel, Prettenhofer, Weiss, Dubourg, Vanderplas, Passos, Cournapeau, Brucher, Perrot, \& Duchesnay}]{scikit-learn}
Pedregosa, F., Varoquaux, G., Gramfort, A., {et~al.} 2011, Journal of Machine Learning Research, 12, 2825

\bibitem[{{Sadykov} \& {Kosovichev}(2017)}]{sadykov2017PIL}
{Sadykov}, V.~M., \& {Kosovichev}, A.~G. 2017, \apj, 849, 148, \dodoi{10.3847/1538-4357/aa9119}

\bibitem[{{Scherrer} {et~al.}(2012){Scherrer}, {Schou}, {Bush}, {Kosovichev}, {Bogart}, {Hoeksema}, {Liu}, {Duvall}, {Zhao}, {Title}, {Schrijver}, {Tarbell}, \& {Tomczyk}}]{2012SoPh..275..207S}
{Scherrer}, P.~H., {Schou}, J., {Bush}, R.~I., {et~al.} 2012, \solphys, 275, 207, \dodoi{10.1007/s11207-011-9834-2}

\bibitem[{Sun {et~al.}(2022)Sun, Bobra, Wang, Wang, Sun, Gombosi, Chen, \& Hero}]{sun2022predicting}
Sun, Z., Bobra, M.~G., Wang, X., {et~al.} 2022, The Astrophysical Journal, 931, 163

\bibitem[{Wang {et~al.}(2020)Wang, Chen, Toth, Manchester, Gombosi, Hero, Jiao, Sun, Jin, \& Liu}]{wang2020predicting}
Wang, X., Chen, Y., Toth, G., {et~al.} 2020, The Astrophysical Journal, 895, 3

\bibitem[{Yeolekar {et~al.}(2021)Yeolekar, Patel, Talla, Puthucode, Ahmadzadeh, Sadykov, \& Angryk}]{9679962}
Yeolekar, A., Patel, S., Talla, S., {et~al.} 2021, in 2021 International Conference on Data Mining Workshops (ICDMW), 1067--1076, \dodoi{10.1109/ICDMW53433.2021.00138}

\bibitem[{Yu {et~al.}(2009)Yu, Huang, Wang, \& Cui}]{yu2009short}
Yu, D., Huang, X., Wang, H., \& Cui, Y. 2009, Solar Physics, 255, 91

\bibitem[{Yuan {et~al.}(2010)Yuan, Shih, Jing, \& Wang}]{yuan2010automated}
Yuan, Y., Shih, F.~Y., Jing, J., \& Wang, H.-M. 2010, Research in Astronomy and Astrophysics, 10, 785

\bibitem[{Zhang {et~al.}(2022)Zhang, Li, Yang, Jing, Wang, Wang, \& Shang}]{zhang2022solar}
Zhang, H., Li, Q., Yang, Y., {et~al.} 2022, The Astrophysical Journal Supplement Series, 263, 28

\end{thebibliography}
\bibliographystyle{aasjournal}



\end{document}